\documentclass[fleqn,usenatbib]{mnras}

\usepackage{epstopdf}
\usepackage{graphicx}
\usepackage{txfonts}

\usepackage{soul}

\newcommand\nh{$N(\mathrm{H}_2)$}

\usepackage{longtable}
\usepackage{rotating}
\usepackage{pdflscape}
\usepackage{multirow}
\usepackage{color}

\defcitealias{2016MNRAS.459..342M}{Paper I} 	  	

\title[A census of dense cores in TMC1]{\textit{Herschel} Gould Belt Survey in Taurus. II: A census of dense cores and filaments in the TMC1 region}
\author[J.M. Kirk et al.]{
{\parbox{\textwidth}{J. M. Kirk$^1$\thanks{E-mail: JMKirk@uclan.ac.uk}, 
D. Ward-Thompson$^1$,
J. Di Francesco$^2$,
Ph. Andr\'e$^3$,
D. W. Bresnahan$^1$,
V. K\"onyves$^1$,
K. Marsh$^{4,5}$,
M. J. Griffin$^4$,
N. Schneider$^6$,
A. Men'shchikov$^3$
P. Palmeirim$^7$,
S. Bontemps$^8$,
D. Arzoumanian$^{9}$
M. Benedettini$^{10}$,
S. Pezzuto$^{10}$,
}}
\\
$^1$Jeremiah Horrocks Institute, University of Central Lancashire, Preston, Lancashire, PR1 2HE, UK. \\
$^2$National Research Council of Canada, Herzberg, Astronomy \& Astrophysics Research Centre, 5071 West Saanich Road, Victoria (BC), V9E 2E7, Canada\\
$^3$Laboratoire Astrophysique (AIM), Universit\'e Paris-Saclay, Universit\'e Paris Cit\'e, CEA,CNRS, AIM, 91191 Gif-sur-Yvette, France\\
$^4$School of Physics and Astronomy, Cardiff University, Queens Buildings, The Parade, Cardiff, Wales, CF24 3AA, UK \\
$^5$IPAC, Caltech, 1200E California Boulevard, Pasadena, CA 91125, USA\\
$^6$Physikalisches Institut, Universit{\"a}t zu K{\"o}ln, Z{\"u}lpicher Str. 77, 50937 K{\"o}ln, Germany\\
$^7$Instituto de Astrof{\'i}sica e Ci\^encias do Espa\c{c}o, Universidade do Porto, CAUP, Rua das Estrelas, P-4150-762 Porto, Portugal\\
$^8$Laboratoire d’astrophysique de Bordeaux, Univ. Bordeaux, CNRS, B18N, all\'{e}e Geoffroy Saint-Hilaire, 33615 Pessac, France\\
$^9$National Astronomical Observatory of Japan, 2-21-1 Osawa, Mitaka, Tokyo 181-8588, Japan \\
$^{10}$INAF – Istituto di Astrofisica e Planetologia Spaziali, via Fosso del Cavaliere 100, 00133 Roma, Italy\\
}

\date{Accepted XXX. Received YYY; in original form ZZZ}

\pubyear{2023}

\begin{document}
\label{firstpage}
\pagerange{\pageref{firstpage}--\pageref{lastpage}}

\maketitle

\begin{abstract}
We present a catalogue of dense cores and filaments in a  $3.8^\circ\times2.4^\circ$ field around the TMC1 region of the Taurus Molecular Cloud. The catalogue was created using photometric data from the \textit{Herschel} SPIRE and PACS instruments in the 70\,$\mu$m, 160\,$\mu$m, 250\,$\mu$m, 350\,$\mu$m, and 500\,$\mu$m continuum bands. Extended structure in the region was reconstructed from a \textit{Herschel} column density map. Power spectra and PDFs of this structure are presented. 
The PDF splits into log-normal and power-law forms, with the high-density power-law component associated primarily with the central part of TMC1. The total mass in the mapped region is 2000~M$_\odot$, of which 34\% is above an extinction of $A_V\sim3$ mag -– a level that appears as a break in the PDF and as the minimum column density at which dense cores are found.  A total of 35 dense filaments were extracted from the column density map. These have a characteristic FWHM width of 0.07~pc, but the TMC1 filament itself has a mean FWHM of $\sim0.13$~pc.
The thermally supercritical filaments in the region are aligned orthogonal to the prevailing magnetic field direction. 
Derived properties for the supercritical TMC1 filament support the scenario of it being relatively young. A catalogue of 44 robust and candidate prestellar cores is created and is assessed to be complete down to 0.1~M$_\odot$. 
The combined prestellar CMF for the TMC1 and L1495 regions is well fit by a single log-normal distribution and is comparable to the standard IMF.
\end{abstract}

\begin{keywords}
ISM: individual objects: TMC1 molecular cloud, -- stars: formation -- ISM: clouds -- ISM structure -- submillimeter: ISM -- catalogues
\end{keywords}

\section{Introduction}

Stars form within the dense cores of giant molecular clouds. Observations with the \textit{Herschel} Space Observatory \citep{2010A&A...518L...1P} have reinforced the paradigm that these cores form preferentially as condensations on filaments within the clouds \citep[e.g.,][]{2014prpl.conf...27A}. The success of \textit{Herschel} in this regard is due to its unparalleled ability to map simultaneously in five submillimetre wavelengths large contiguous areas of sky with unprecedented dynamic range and sensitivity. The  \textit{Herschel} Gould Belt Survey \citep[HGBS;][]{2010A&A...518L.102A} has used this ability to map systematically star formation regions in the Gould Belt \citep[a $\sim 1$~kpc loop of molecular clouds that encircle the local bubble, see][]{1847herschel,1879gould}. Catalogues and analysis of dense cores in Aquila \citep{2015A&A...584A..91K}, Corona Australis \citep{2018arXiv180107805B}, and Lupus \citep{2018A&A...619A..52B} have already been published.

Taurus forms broadly two large parallel structures which have been mapped with \textit{Herschel} \citep{2013MNRAS.432.1424K}. The overall mapping approach is described in \citet{2013MNRAS.432.1424K}. The first part of the Taurus catalogue was published in \citet[][hereafter {\bf Paper I}]{2016MNRAS.459..342M}.  \citetalias{2016MNRAS.459..342M} found a total of 52 prestellar cores out of a large population of 525 unbound starless cores across an area of $\sim 4^\circ \times 2^\circ$. They showed that the completeness-corrected mass function for the unbound starless cores was consistent with a power-law form below a mass of $0.3$~M$_\odot$. This correspondence was unlike the mass function for bound prestellar cores which followed the classical log-normal form, in agreement with previous studies.     

The area mapped in \citetalias{2016MNRAS.459..342M} was the L1495 region. Its dominant feature is the B211/213 filament which stretches across $\sim2$ degrees ($\sim5$~pc at 140~pc). \citet{2013A&A...550A..38P} showed that the B211/213 was an archetypical case of material being funnelled along low-density striations onto the central filament. For example, the striations and filament were respectively oriented parallel and perpendicular to the local magnetic field direction. The next HGBS tile across from the L1495 region, along the northern structure, is the one presented in this paper. It is markedly different from the B211/213 and L1495 region as it lacks any dominant singular filament and is instead composed of a network of distributed filaments and structures (see Figure~\ref{fig:nh2} and Section~\ref{sec:structure}). For ease of reference, we refer to this entire region, i.e., the area mapped in this paper, as the ``TMC1 region.''

This paper is laid out as follows. Section~\ref{sec:obs} describes the \textit{Herschel} observations of the TMC1 region and the data reduction process. Section~\ref{sec:dist} characterises the distance to the TMC1 region. Section~\ref{sec:structure} describes the creation of a column density map from the \textit{Herschel} observations and includes an analysis of its power spectrum and hierarchical structure. In Section~\ref{sec:filaments}, we characterise the filamentary structure within the TMC1 region and compare it to the bulk magnetic field and the location of the cores in our catalogue. In Section~\ref{sec:catalog} we describe our source extraction methodology, the resulting catalogue of cores, and the catalogue's general properties, and the properties of the prestellar core mass function in Taurus.

\section{Observations}
\label{sec:obs}

\begin{table}
\caption{\label{tab:filters} Filter-dependent observation details. Columns 1 and 2 list the filter name and wavelength. Column 2 lists the assumed calibration uncertainty. Column 4 lists the effective HPBW of the telescope at this wavelength.  Column 5 lists the median {\it Planck} DC level across the map (see Section~\ref{sec:structure}).}
\centering{
\begin{tabular}{lrccc}
\hline
Filter	& $\lambda$ & $\sigma_{\mathrm{cal}}$ & HPBW & $S_\mathrm{Planck}$ \\
	& [$\mu$m] & [\%] & [''] & [MJy/sr]  \\
\hline
PACS blue	&	70	& 10 & 8.4 &  4.62   \\
PACS red	& 	160	& 10 & 14 &  77.4 \\
SPIRE PSW	&	250	& 12 & 18.2 & 59.4  \\
SPIRE PMW	&	350	& 12 & 25 &  36.9 \\
SPIRE PLW	&	500	& 12 & 36 &  17.1 \\
\hline
\end{tabular}
}
\end{table}

\begin{table*}
\caption{\label{tab:aors} Taurus Molecular Cloud observation details. Column 1 lists the tile name. Columns 2 and 3 list the Right Ascension and Declination (J2000) of the field centre. Column 4 lists the observations' AOR identification number. Column 5 lists the OD, the day of the observation numbered from the launch of the spacecraft, and column 6 lists the UT date of the observation. Column 7 lists the scan direction of the field -- N for nominal, O for an orthogonal cross-scan.}
\centering{
	\begin{tabular}{l r@{$^{h}$}c@{$^{m}$}c@{$^{s}$\hspace{2mm}} r@{\degr}c@{\arcmin}c@{\arcsec\hspace{2mm}} rrrrr }
	\hline
	Field & \multicolumn{3}{c}{R.A.} & \multicolumn{3}{c}{Dec.} & AOR & Duration & OD & UT Date & Scan Dir \\
	\hline
	N2 & 4 & 37 & 31.1 & +26 & 01 & 22 & 1342202252 & 10086 & 451 & 2010-08-07 & N \\
	   & 4 & 37 & 31.1 & +26 & 01 & 22 & 1342202253 & 14453 & 451 & 2010-08-07 & O \\
	\hline
	\end{tabular}
}
\end{table*}

The \textit{Herschel} Space Observatory was a 3.5m-diameter passively-cooled Cassegrain space telescope launched on the 14th of May, 2009 \citep{2010A&A...518L...1P}. It operated two photometric instruments in the far-infrared and submillimetre regimes---the Photodetector Array Camera and Spectrometer \citep[PACS, ][]{2010A&A...518L...2P} and the Spectral and Photometric Imaging Receiver, \citep[SPIRE, ][]{2010A&A...518L...3G,2010A&A...518L...4S}---to provide sufficient wavelength coverage to bracket the peak of the 10\,K cold dust SED. Table~\ref{tab:filters} lists the instrument and filter names, the wavelengths of the filters, and our assumed calibration errors.

The HGBS\footnote{\url{http://gouldbelt-herschel.cea.fr}} \citep{2010A&A...518L.102A} mapped the Taurus Molecular Cloud (TMC) using its SPIRE/PACS parallel-scanning mode. The mapping scheme is described in \citet{2013MNRAS.432.1424K}, this paper presents data from Tile N2 of that scheme.
The observation details for this tile are listed in Table \ref{tab:aors}. The parallel mode observes simultaneously with all three SPIRE filters and two of the three PACS filters, and the scanning speed was 60 arcsec/s. The tile was observed twice, in perpendicular mapping directions, to aid in the removal of $1/f$ noise. The TMC1 observations used the 70/160\,$\mu$m\, PACS filter set. 
In parallel-mode, the effective PACS HPBW at 160\,$\mu$m was degraded due to the elongation of the PSF in the scan direction caused by sampling limitations imposed by the SPIRE/PACS Parallel Mode. 
The resulting telescope HPBWs are also listed in Table~\ref{tab:filters}.

The SPIRE data were reduced to level 1 status (flux calibrated timelines) using HIPE version 10 \citep{2010ASPC..434..139O}. The SPIRE thermistor timelines were visually inspected for discontinuities caused by instrumental factors. These were corrected before the timelines were processed.
The data were de-glitched using the concurrent and wavelet deglitcher tasks. Residual glitches were identified in restored maps and manually flagged as bad in the timeline data. Flux calibration, temperature drift correction, and bolometer time response corrections were done with the standard SPIRE pipeline tasks using version 10.1 of the SPIRE calibration product.
The level 1 timelines for each contiguous region were merged into a single level 1 product. The differences in relative bolometer-to-bolometer gain were corrected using SPIRE calibration product version 10.1. Residual  $1/f$ noise in the observation baselines was removed by using the destriper task in HIPE.
The final SPIRE maps were made using the naive map maker and were produced on pixel grids of 6\arcsec, 10\arcsec, and 14\arcsec\ pixels for the 250-, 350-, and 500-$\mu$m maps, respectively.

The PACS data were reduced to level 1 status using HIPE version 10.0 (build 2843) with the standard PACS pipeline. Bad and saturated pixels were masked during the reduction process, glitches were removed following the procedure described in \citet{2015A&A...584A..91K}, and version PACS\textunderscore CAL\textunderscore 69\textunderscore 0 of the PACS calibration product was applied. For map making, the level 1 timelines were exported to the IDL Scanamorphos map maker \citep[version 20.0;][]{2013PASP..125.1126R} which was used to remove thermal and non-thermal brightness drifts and any residual glitches. The resulting PACS maps were produced on a 3\arcsec/pixel grid.

\section{TMC1 Distance}
\label{sec:dist}
The Taurus Molecular Cloud is one of the nearest star formation regions to the Sun. Tomographic analysis of the GAIA distances and tracers of molecular material has shown that the full Taurus molecular cloud is a U-shaped structure that points away from us and the local bubble \citep{2021A&A...652A..22I,2022A&A...658A.166D}. The depth of the material may be 20~pc or more; the TMC1 region is on the leading edge of that structure \citep{2021A&A...652A..22I}. 

The traditional distance to the Taurus region is usually taken as 140 pc \citep{1980AcA....30..541S,1994AJ....108.1872K,2008loinard}. However, the launch of the GAIA parallax mission has allowed newer studies to probe the distance and three-dimensional structure of the cloud \citep{2018AJ....156..271L,2019arXiv190406980F,2019A&A...630A.137G}. 

There is an extended YSO population across Taurus \citep[e.g.][]{2019AJ....158...54E}. Distances are attributed to parts of the TMC by grouping GAIA sources into clusters/groups and then associating the mean distance of that cluster to part of the cloud. If a single cluster is associated with the entire region, the mean distance assigned is 141-142~pc \citep{2018AJ....156..271L,2019AJ....158...54E,2021AJ....162..110K}. Splitting those stars into two cluster groups gives distances $138\pm1.9$ and $142\pm2$ \citep{2019A&A...630A.137G}. 

\begin{figure}
    \centering
    \includegraphics[width=\columnwidth]{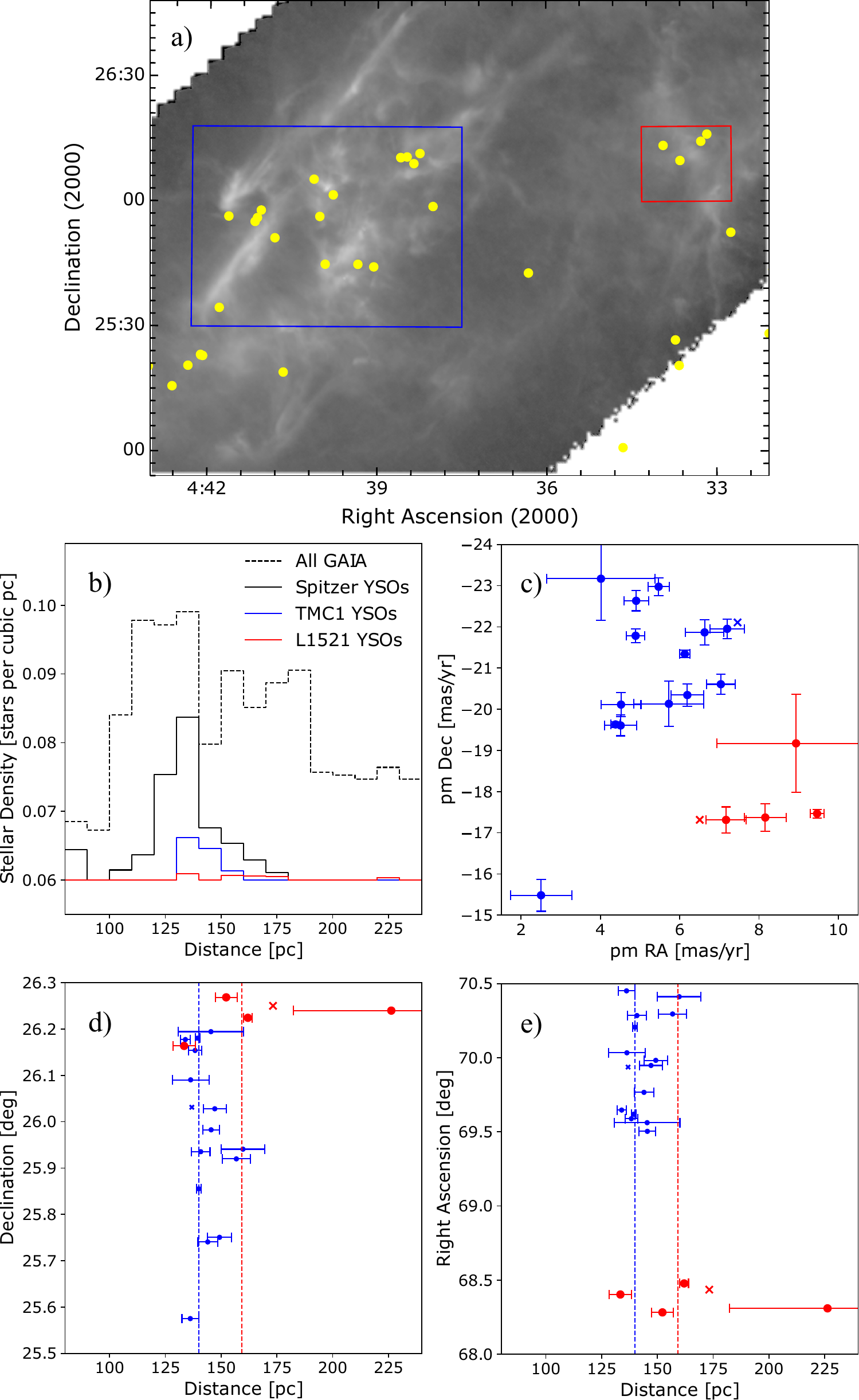}
    \caption{The distribution of \textit{Spitzer} detected YSOs with GAIA DR3 distances towards the TMC1 region. 
    \textbf{(a)} The positions of Spitzer detected YSO candidates \citep[yellow markers;][]{2010ApJS..186..259R} with a \textit{GAIA} distance in our field. The grayscale is the \nh\ column density map as per Figure~\ref{fig:nh2}. The blue box denotes those YSOs associated with TMC1 while the red box shows those associated with L1521. Sources within those boxes are colour-coded blue (TMC1) and red (L1521) in the other plots in this Figure. \textbf{(b)} The number volume density of GAIA sources across the field studied in this paper. The solid black line shows all GAIA sources, the dash black line shows all \textit{Spitzer} YSOs, the blue curve show YSOs associated with TMC1, and the red curve shows YSOs associated with L1521. \textbf{(c)}, \textbf{(d)}, and \textbf{(e)} show the distribution of the TMC1 (blue) and L1521 (red) YSOs in parameter space. The dashed lines show the weighted average distances to TMC1 (blue) of $140.0\pm0.6$~pc and L1521 (red) of $159\pm2$~pc.}
    \label{fig:gaia}
\end{figure}

These studies use catalogues of YSOs with a range of ages, some of which may be  more evolved and thus dissociated from the material from which they formed. We double-check the distance attribution by selecting only those GAIA sources with a \textit{Spitzer} detection from \citet{2010ApJS..186..259R}. Infrared-detected sources should be the younger subset of young stars in the region and, thus, those most likely to be still associated with their natal cloud. Figure~\ref{fig:gaia}a shows the location of YSOs and YSO candidates reported by \citet{2010ApJS..186..259R}. Two groups of YSOs towards TMC1 and L1521 are indicated. Figure~\ref{fig:gaia}b shows that the distance trend for all GAIA sources in this field is to peak at 140~pc and drop away at higher distances, the assumption being that sources behind the Taurus cloud are extincted by it. The YSOs towards this part of Taurus also peak at 140~pc, albeit with a short tail at higher distances. Figure~\ref{fig:gaia}c to e show that the TMC1 and L1521 sources are well separated in proper motion. However, the L1521 YSOs appear to occupy a spread of distances with a mean of $159\pm2$~pc. Sources towards TMC1 are more tightly clustered with a mean distance of $140.0\pm0.6$.

Therefore despite different clustering techniques, sample preferences, and methods - the underlying distance for the TMC1 region remains resolutely stuck at its historical value of 140~pc. It is a value that may no longer be as useful for the wider region but has been shown to repeatedly hold as the preferred distance to TMC1/HCL2/L1521. Therefore, we adopt 140~pc as the distance to structures studied in this paper. The caveat is that 140~pc is the mean distance to the young stellar sources associated with the cloud; their projected scatter \citep{2019A&A...630A.137G,2019AJ....158...54E,2021AJ....162..110K} and the width of the cloud Figure~\ref{fig:gaia} is on the order of a degree. At 140~pc this equates to a plane-of-the-sky distance of 2.4~pc.

\section{Extended Structure}
\label{sec:structure}

Figures~\ref{fig:flux070}--\ref{fig:flux500} show the five \textit{Herschel} flux density maps towards the TMC1 region. There is a change in the morphology of the emission with wavelength. The 70-$\mu$m map only shows compact thermal emission from young protostars. The 160-$\mu$m map shows more extended structure, with the 160- and 250-$\mu$m maps showing significant levels of cirrus-like clouds. At 350 and 500$\mu$m, the maps show thick, colder filaments and dense cores begin to stand out. To analyse star formation within the TMC1 region, we need to be able to extract the most appropriate structures from this assemblage. We assume a single isothermal dust temperature along the line of sight. 
In this Section, we use a \nh\ column density map to examine the extended structures associated with the colder dust, and in the following Section, we look at the properties of cores extracted from these five maps.

\subsection{Column Density Maps}
\label{coldens}

\begin{figure*}
\centering{
    \includegraphics[width=0.95\textwidth]{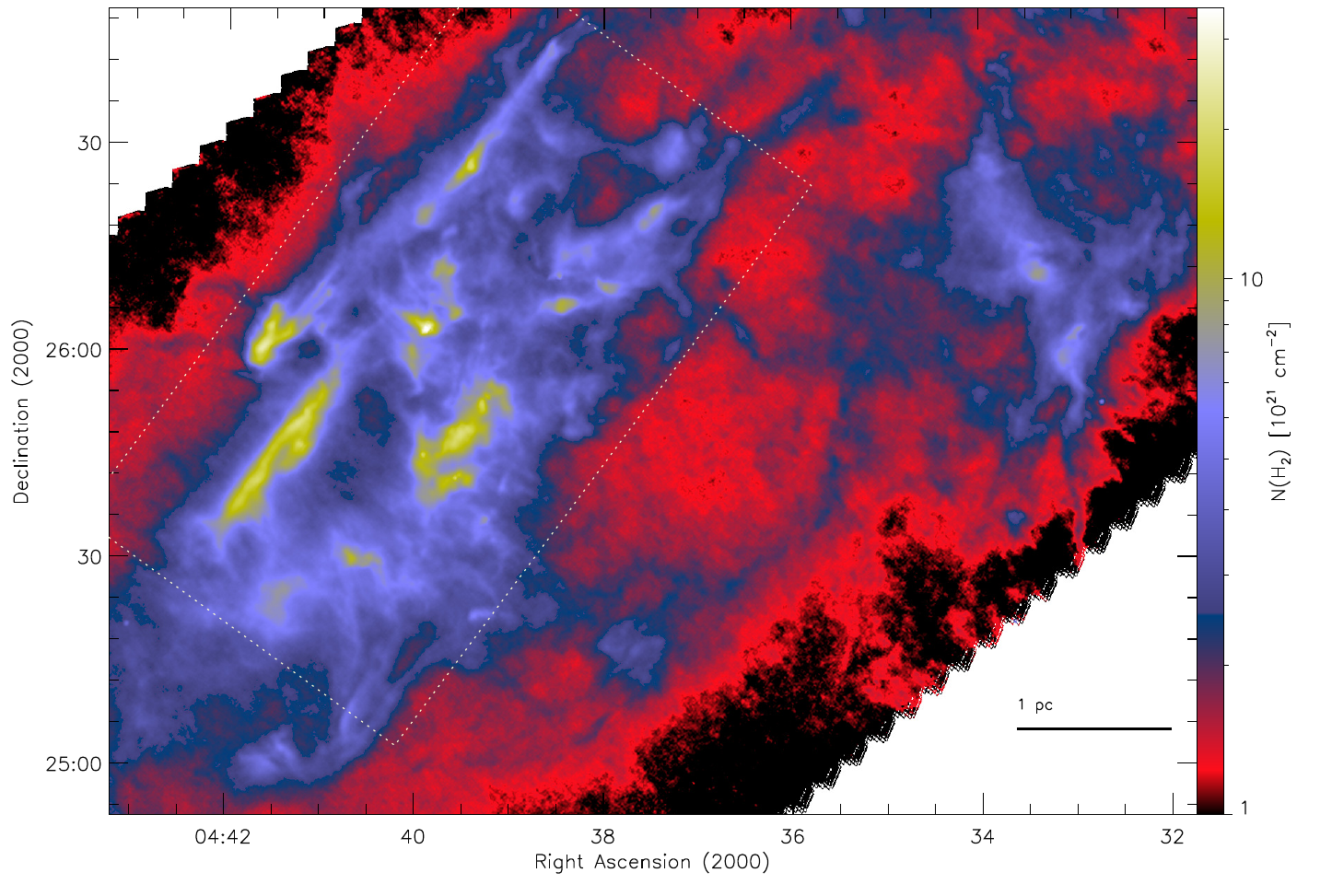} \\
    \includegraphics[width=0.95\textwidth]{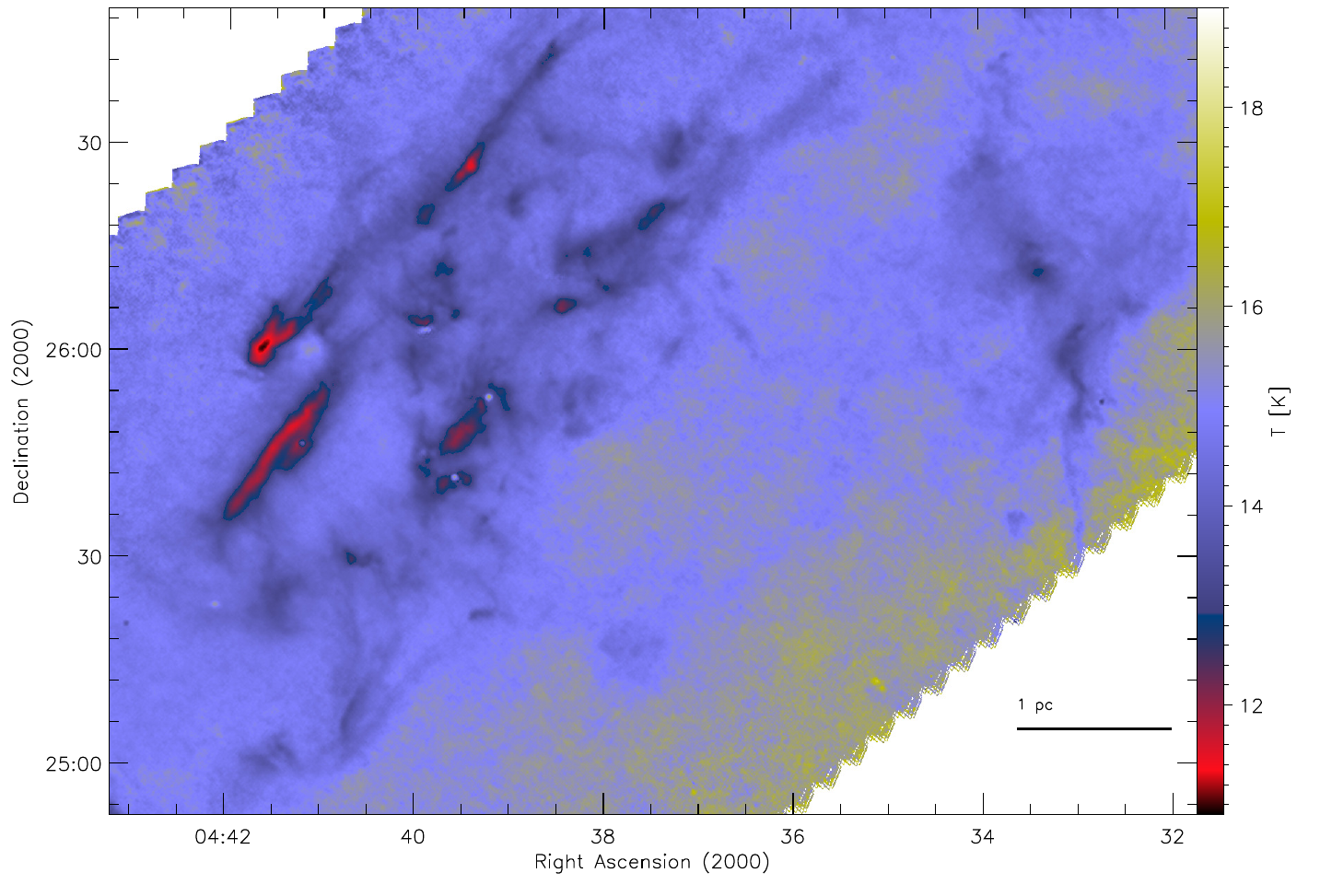}
}
\caption{\label{fig:nh2} Maps of \nh\ column density ({\bf top}) and dust temperature ({\bf bottom}) towards the TMC1 region derived from \textit{Herschel} 160-$\mu$m to 500-$\mu$m data. The colour bars show the range of the displayed quantities. The effective resolution of the maps is 36\arcsec. The horizontal yardstick shows 1 pc at the assumed distance to Taurus (140 pc). The dotted-box in the upper figure shows the region used to determine the PDFs in Section~\ref{sec:pdf} and Figure~\ref{fig:pdf}.}
\end{figure*}

An $N(\mathrm{H}_2)$ column density map was created on the 250-$\mu$m pixel grid and at the 500-$\mu$m resolution using the method described in \citet{2013MNRAS.432.1424K}. In summary, the median DC level at each wavelength was bootstrapped to the median flux estimated from {\it Planck} \& IRAS \citep[cf.][ see Table~\ref{tab:filters} for the estimated levels]{2010A&A...518L..88B} and the data were convolved to the SPIRE 500-$\mu$m beam psf using the convolution kernels of \citet{2011PASP..123.1218A}.  The function $F_\nu = M B_\nu(T) \kappa_\nu / D^2$ was then fit to the SED of each pixel from 160\,$\mu$m to 500\,$\mu$m where $F_\nu$ is the monochromatic intensity, $M$ is the dust mass, $B_\nu(T)$ is the Planck function at temperature $T$, and $D$ is the average assumed distance to the TMC1 region (140~pc). The dust mass opacity $\kappa_\nu$ was parameterized as $\kappa_\nu\propto\nu^2$ (i.e., assuming $\beta=2$) and was referenced against a value of 0.1\,cm$^2$\,g$^{-1}$ at 300 $\mu$m  \citep{1983QJRAS..24..267H}. In converting dust mas to molecular gas mass we assumed a dust-to-molecular hydrogen mass ratio of 100. The minimisation package MPFIT \citep{2009ASPC..411..251M} was used for the fitting and the per-pixel flux errors included contributions from the map rms, DC level estimation, and calibration uncertainty (for details, see \citealt{2013MNRAS.432.1424K}).

Figure~\ref{fig:nh2} (top panel) shows the $N(\mathrm{H}_2)$ map at 36\arcsec\ resolution. The field comprises a contiguous region (in the plane of the sky) of high column density material in the north-eastern (left) part of the field. This feature is Heiles Cloud 2 \citep[HCL2;][]{1968ApJ...151..919H}. HCL2 was one of the dark clouds where normal OH emission was first detected \citep{1968ApJ...151..919H}. The hole at approximately 4:40:45, +25:45:00 has led to the features surrounding it being dubbed the ``Taurus Molecular Ring,'' albeit one that breaks up into individual dense cores when viewed in more detail \citep{2004A&A...420..533T}. The most prominent part of this ring is Taurus Molecular Cloud No. 1 (TMC1), a $\sim0.5^\circ$ long linear feature on its north-east side. The western feature (right) is the original position of the original L1521 dark nebula \citep{1962ApJS....7....1L}.

Figure~\ref{fig:nh2} also shows the temperature map also derived from the modified blackbody SED fitting. Comparing the Molecular Ring column density structure to the dust temperature map (lower panel) shows that the parts with the highest column densities are associated with cold dust. In those parts, the temperatures drops to 11-12\,K, down from the general $\sim14$\,K found over most of the region. The \textit{Herschel} emission is a product of different temperature elements along the line of sight, so the dust temperature produced from it will be a characteristic average. 

TMC1 was studied in IRAS and SCUBA dust emission by \citet{2008MNRAS.384..755N} who dubbed it the ``Bull's Tail.'' After background subtraction, they found that the filament was consistent with a narrow, cold ($\sim8$~K) core and a broader, warmer ($\sim12$~K) jacket. The Bull's Tail is visible as a cool feature (the red strip in the bottom panel of Figure~\ref{fig:nh2}). Indeed, this filament was sufficiently dense to show in absorption in a \textit{Spitzer}~70-$\mu$m map \citep{2008MNRAS.384..755N}. Following \citet{2021A&A...645A..55P}, the SED fit used to produce the column density map was used to create an expected map of 70-$\mu$m intensity. The expected map showed absorption in places of high-column density, something not seen in the original PACS 70-$\mu$m intensity map. \citet{2021A&A...645A..55P} found the same pattern in Perseus and ruled out errors in the fitting of the dust temperature as a cause of the discrepancy of the actual and expected 70-$\mu$m intensity. That this extinction is seen with \textit{Spitzer}, but not \textit{Hershel} could imply a methodological cause (i.e., the way the data was processed in either study). 

As mentioned previously, the structure visible in Figure~\ref{fig:nh2} is markedly different than that seen in the neighbouring L1495 region (cf. fig. 3 of \citealt{2013A&A...550A..38P} or fig. 1 of \citealt{2016MNRAS.459..342M}) which is dominated by a single strong trunk filament. By contrast, the TMC1 region is more complex and nest-like, albeit with some fainter but strongly linear features of its own (see Section~\ref{sec:filaments}). 
This is similar to the ridge-nest structures observed in the Vela C region \citep{2011A&A...533A..94H}, albeit on a smaller physical scale. 

\subsubsection{Comparison with previous maps}

\begin{figure*}
\includegraphics[width=0.33\textwidth]{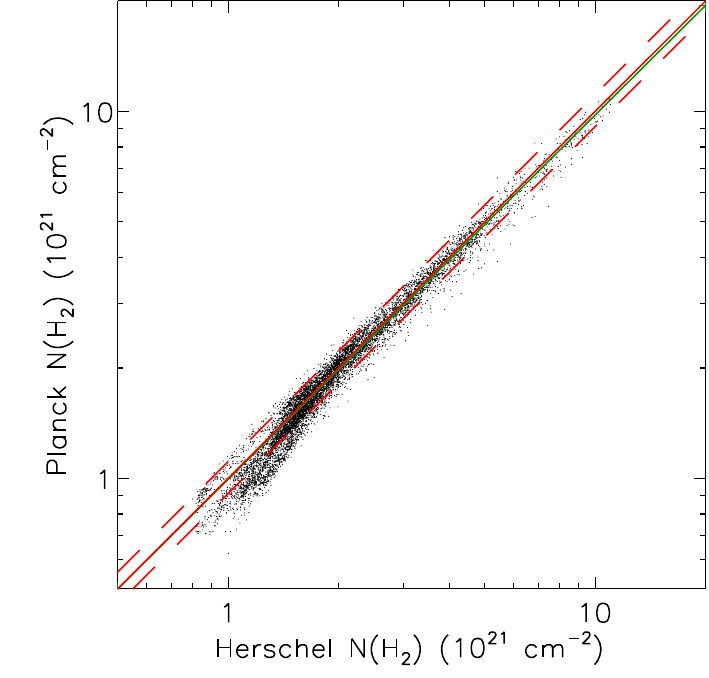}
\includegraphics[width=0.33\textwidth]{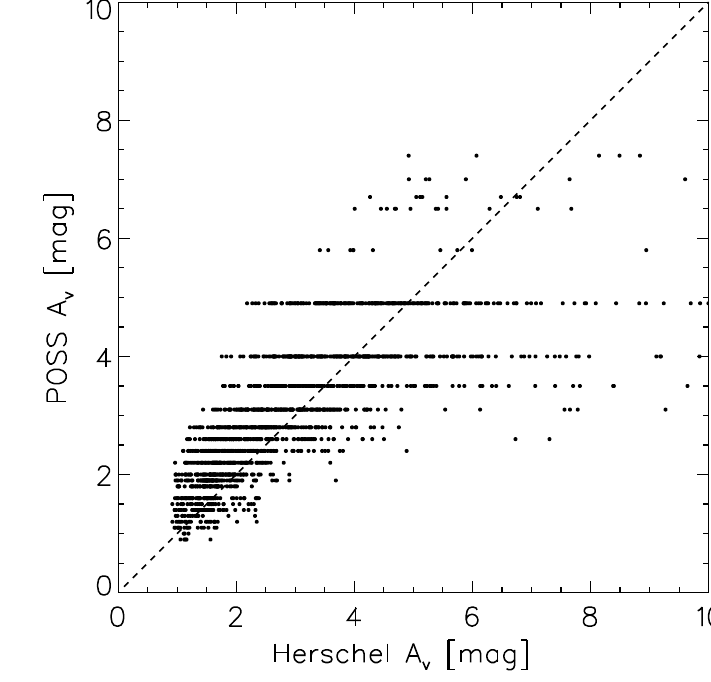}
\includegraphics[width=0.33\textwidth]{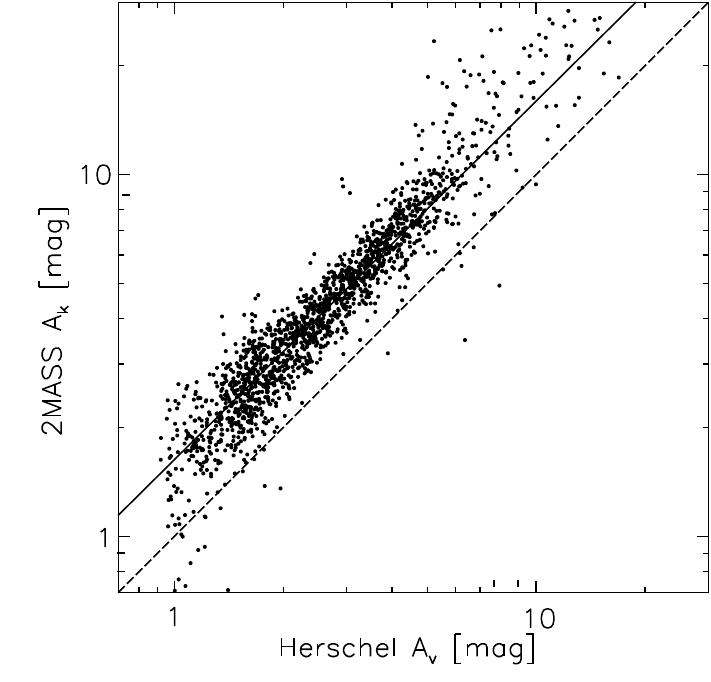}
\caption{\label{fig:planck} Point-by-point comparisons of the \textit{Herschel} column density values (or equivalent visual extinctions) and equivalent datasets. \textbf{(Left)} A position-by-position comparison of the \textit{Herschel} column densities smoothed to a 5\arcmin\ HPBW versus an equivalent column density calculated from the {\it Planck} 353 GHz optical depth. The two dashed-red lines show a deviation of 10\% away from a strict 1:1 relationship (the solid-red line). A best-fit power-law relation (fitted above $1.5\times10^{21}$~cm$^{-2}$) between the two is shown by a green line and has the form $N(\mathrm{H}_2)_{Planck} = 0.994\pm0.002 \times N(\mathrm{H}_2)_{Herschel}^{ 0.993\pm0.002 }$.  \textbf{(Center)} A comparison of $A_V$ calculated from the \textit{Herschel} column densities versus the $A_V$ estimated from the POSS plates towards HCL2 \citep{1984A&AS...58..327C}. The dashed-line is a line of equivalence. \textbf{(Right)} A comparison of $A_V$ calculated from the \textit{Herschel} column densities versus the K-band extinction calculated from the 2MASS catalogue catalogue \citep{2011A&A...529A...1S}. The dashed-line is a line of equivalence. The solid-line is linear fit to the data with the form $A_K = 1.63\ Av^{0.99\pm0.06}$. }
\end{figure*}

We check the absolute calibration of the $N(\mathrm{H}_2)$ map by using the {\it Planck} 353 GHz (850 $\mu$m) dust optical-depth map of the region \citep{2014A&A...571A..11P}. The \textit{Herschel} N(H$_2$) map was smoothed to the same resolution as the {\it Planck} map (5\arcmin\ HPBW). A {\it Planck} column density map was then extrapolated from their published optical depth and temperature maps using the same reference properties used for the \textit{Herschel} map. Both column density maps were then regridded to a common pixel grid. Figure~\ref{fig:planck} (left) shows a pixel-by-pixel comparison of the column densities from the two maps. The horizontal axis shows the \textit{Herschel} N(H$_2$) column density at 5\arcmin\ while the vertical axis shows the extrapolated {\it Planck} column density. The solid red line shows a 1:1 correlation and the two dashed-lines show a 10\% variance from that in either property.

Figure~\ref{fig:planck} (left) shows a small dip away from equivalence below ${\sim}1.5\times10^{21}$~cm$^{-2}$. This deviation could be attributed to warmer, low-density material to which \textit{Herschel} is more sensitive than \textit{Planck} (due to its coverage at shorter wavelengths). A linear fit to the data above this point gives an excellent correlation to the data (shown by the green line). The two results show that the offset-correction method from \citet{2010A&A...518L..88B} has well reproduced the absolute level of emission in the map.

The \textit{Herschel} column densities can be converted into an equivalent visual extinction using standard assumptions \citep{1978ApJ...224..132B}. The central panel of Figure~\ref{fig:planck} shows a comparison between these values and the visual extinction values towards HCL2 calculated from the POSS plates by \citet{1984A&AS...58..327C}. The values show good agreement over the range of 1--5 mag. A similar comparison was made to the $A_K$ extinctions calculated from 2MASS sources \citep{2011A&A...529A...1S} and is shown in Figure~\ref{fig:planck} (right). A regression over the data binned in logarithmic bins shows a linear relation of $A_K \propto A_V^{0.99\pm0.06}$ for the TMC1 region. This equivalence differs from that seen in Orion A where a non-linear relationship was found and interpreted as evidence for grain growth \citep{2013ApJ...763...55R}. Likewise, \citet{2021ApJ...915...74U} interpreted a larger average dust grain size in Orion A, compared to Taurus, as evidence of grain growth. Therefore, the linearity of the data in this work would be consistent with the reported differences between Taurus and Orion A.

\subsubsection{High Resolution Column Density Map}

A second ``High Resolution'' column density map was created following the method described in Appendix A of \citet{2013A&A...550A..38P}. This method works by combining the unique features from each of three column density maps with  36\arcsec, 24\arcsec, and 18\arcsec\ resolution (equivalent to 500-$\mu$m, 350-$\mu$m, and 250-$\mu$m HPBW respectively). The 36\arcsec\ map is the same as the map described above. The 24\arcsec\ map was created using the same technique as the 36\arcsec\ map, but was made at the 350-$\mu$m resolution and did not include the 500-$\mu$m data point in the fitting. A colour-temperature map was created from the ratio of the 160-$\mu$m/250-$\mu$m data. For an assumed set of dust parameters -- same as above -- the 160-$\mu$m/250-$\mu$m ratio only depends on the dust temperature. The colour temperature estimated from the 160-$\mu$m/250-$\mu$m ratio was then used with the 250-$\mu$m intensity to calculate a dust mass following the same assumptions as for the regular fitting.

Line-of-sight effects mean that the ``high-resolution column density'' method will underestimate the true column density (as will any per-pixel fitting of absolute fluxes) and combining the different resolutions does increase the noise in the final map \citep{2013A&A...553A.113J,2013A&A...557A..73J}. Nevertheless, the ability to obtain an estimate on these scales outweighs these factors and gives us an essential tool for enhancing our extraction of compact objects.

\subsection{Total Mass In The Region}

\begin{figure}
\includegraphics[width=\columnwidth]{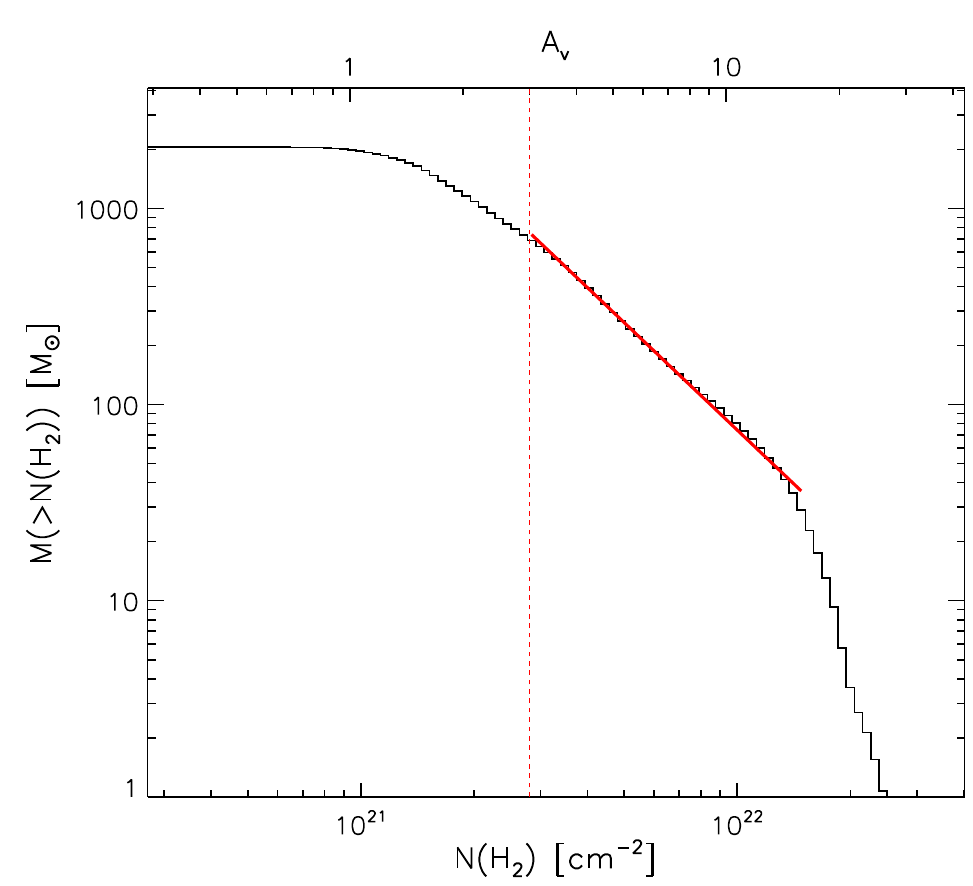} \\
\caption{\label{fig:mass} A cumulative histogram of mass in the 36\arcsec\  resolution $N(\textrm{H}_2)$ map showing the total amount of mass above a given column density threshold. The equivalent optical extinction $A_V$ is shown by the top axis. The red-line gives a best fit power-law index of $-1.82\pm0.02$. The vertical dashed-red line shows the break at $A_V=3$~mag.
}
\end{figure}

Figure~\ref{fig:mass} is a cumulative mass histogram showing the total mass in the region above a given column density threshold. The total mass sampled in the map is $2000$~M$_\odot$. It can be seen that there is a break in the slope of the histogram below $A_V=3$~mag (shown by the vertical dashed-red line). Above this break, the trend can be best-fit with a power-law of $\log M[<N(\mathrm{H}_2)] \propto (-1.82\pm0.02) \log \mathrm{H}_2$. Its index of $-1.82$ is comparable to that of $-1.9$ found in Aquila \citep{2015A&A...584A..91K}. The mass above $A_V=3$~mag is 690~M$_\odot$ (34 per cent of the region total). For comparison, the mass above $A_V=7$~mag, a value sometimes taken as a threshold for star formation \citep{2010ApJ...724..687L}, is 160~M$_\odot$ (${\sim}8$ per cent of region total).

\subsection{Power Spectrum}

\begin{figure}
\includegraphics[width=\columnwidth]{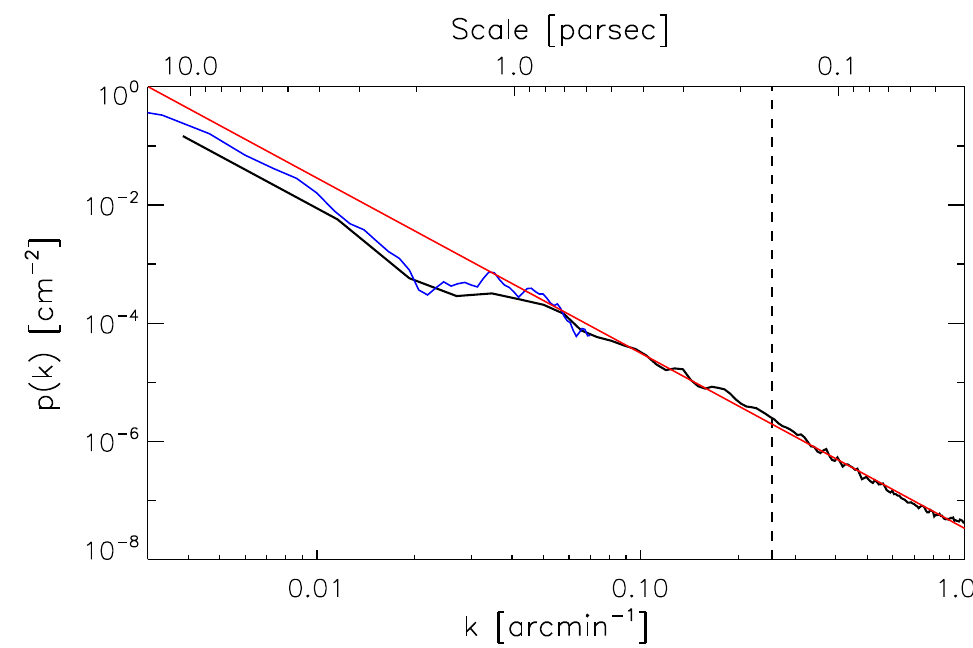} \\
\includegraphics[width=\columnwidth]{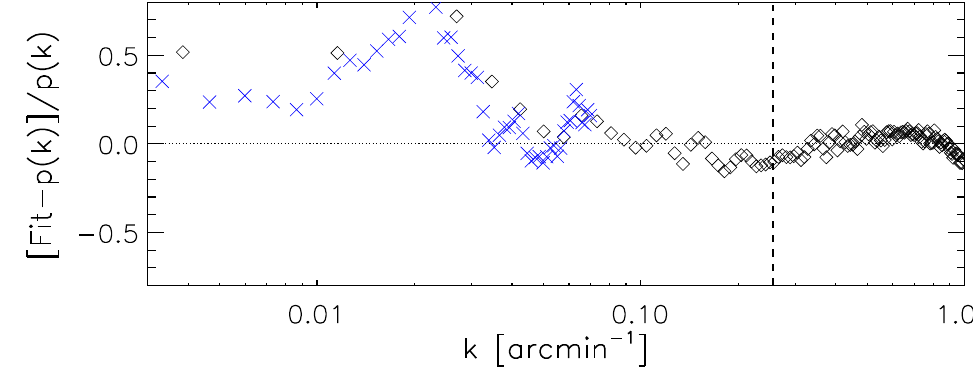}
\caption{\label{fig:powerspec} ({\bf Top}) The $N(\mathrm{H}_2)$ power spectrum  towards the TMC1 region. The horizontal axis shows the wave number $k$ and the vertical axis shows the amplitude of the spectrum $p(k)$. The black \textit{Herschel} $N(\mathrm{H}_2)$ spectrum has been corrected for the instrumental beam and the level of noise in the field \citep{2010A&A...518L.104M}. The red line is a power-law fit to the data on scales smaller than 1~pc (see text for details). The blue curve shows an equivalent power spectrum over the same area calculated from the {\it Planck} $N(\mathrm{H}_2)$ data and assuming a Gaussian 5\arcmin\ beam HPBW. ({\bf Bottom}) The residuals of $p(k)$ minus the best-fit power-law. \textit{Herschel} residuals in black; {\it Planck} residuals in blue. The vertical dashed line shows the equivalent frequency where features of size $0.1$~pc should appear if they are of sufficient amplitude \citep{2017MNRAS.466.2529P,2019A&A...626A..76R}.  }
\end{figure}

Figure~\ref{fig:powerspec} shows the power spectrum of the 500-$\mu$m resolution \textit{Herschel} $N(\mathrm{H}_2)$ column density map (solid black line). This power spectrum been corrected for the level of the noise power spectrum (by subtraction) and the \textit{Herschel} beam \citep[see ][for more details]{2010A&A...518L.104M}. \citet{2010A&A...513A..67B} identified a break at $\sim$1~pc in the power spectrum towards Taurus, which they associated with the anisotropy of Taurus on large scales. Given this break, we fit the power spectrum shown in Figure~\ref{fig:powerspec} on scales beneath it with a power-law of the form $P(k)\propto k^\gamma$ where the best fit value of $\gamma$ is $-2.97\pm0.02$. This value broadly agrees with the value of $-3.1$ found by \citet{2010A&A...513A..67B} over similar scales. 

Also shown in Figure~\ref{fig:powerspec} is a power spectrum of the {\it Planck} $N(\mathrm{H}_2)$ column density (blue curve) over the same area as the \textit{Herschel} data. The spectrum has been corrected for the noise-level on the map and the instrumental beam (assumed to be a 5\arcmin~HPBW Gaussian). This power spectrum closely matches that from \textit{Herschel}, including the dip away from the best-fit power-law on scales below $k{\sim}0.01\textrm{ arcmin}^{-1}$ (above 1~pc). The fact that this dip is seen in both spectra indicates that the cause is something common to both of them.

Analysis of \textit{Herschel} observations shows that filaments in star formation regions have a narrow range of widths, centred on a characteristic mean width of 0.1~pc  \citep{2011A&A...529L...6A,2019A&A...621A..42A,2022A&A...667L...1A}. 
We detect no break in the \textit{Herschel} power spectrum on scales comparable to with scale. The lack of break agrees with the work of \citet{2019A&A...626A..76R} who showed that even if it is present the amplitude of these filaments is too low to be detected.

\subsection{Column Density PDF}
\label{sec:pdf}

\begin{figure}
\includegraphics[width=\columnwidth]{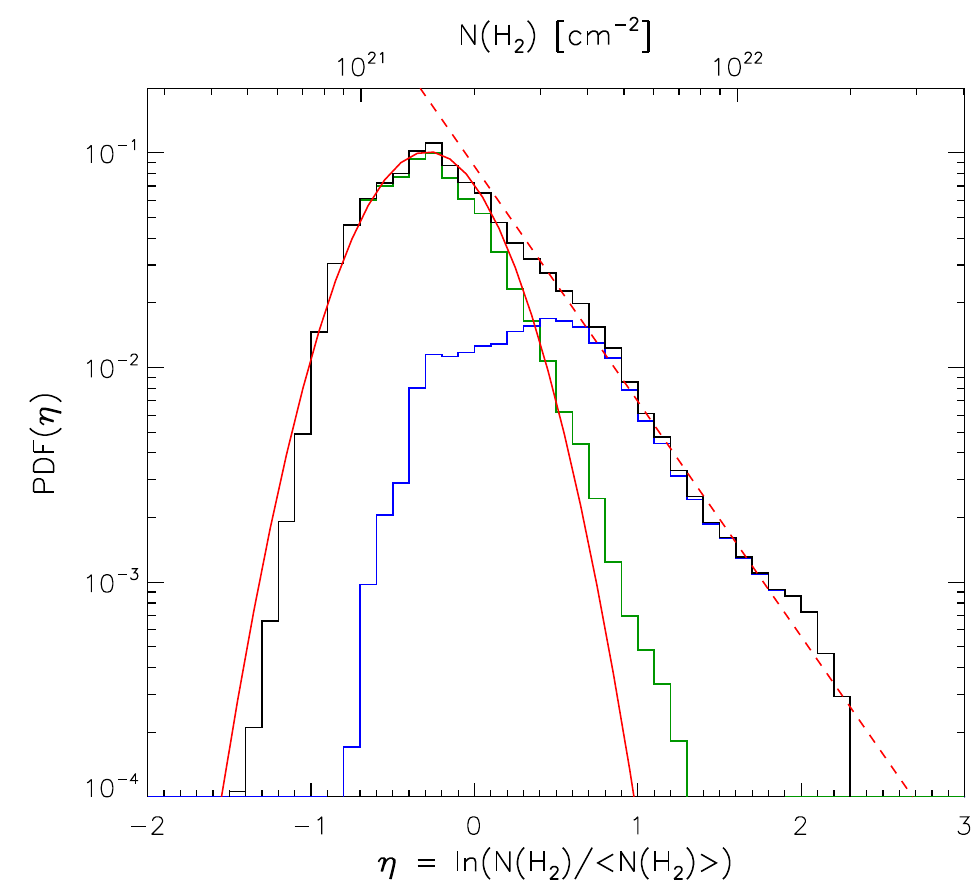} \\
\caption{\label{fig:pdf} PDF (solid-black line) calculated from the \textit{Herschel} column density map of the TMC1 region. The solid-red curve shows a log-normal fit to the peak of the data with a best fit width of $\sigma=0.34$. The dashed-red line shows a power-law fit to the high-density part of the PDF with a best-fit index of $s=-2.48$. The solid-blue line shows the PDF of the TMC1 region alone (shown by the white box in Figure~\ref{fig:nh2}). The solid-green line shows the PDF of the region outside of this box. }
\end{figure}

The Probability Density Function (PDF) of column density in the TMC1 region studied in this paper is plotted in Figure~\ref{fig:pdf} as the solid black line. This effectively shows the number of sight-lines (pixels) through the TMC1 cloud with specific column densities (parameterised as the log of the mean column density). The area under the PDF is normalised to one. The plot shows the characteristic log-normal distribution with a power-law tail at higher column densities \citep{2013ApJ...766L..17S,2015A&A...575A..79S,2015A&A...576L...1L}. To characterise this shape, we fit the log-normal distribution below $\eta\leq 0.2$ with the form
\begin{equation}
PDF(\eta) = \frac{1}{\sqrt{2\pi\sigma^2}}\exp\left(\frac{-(\eta-\eta_0)}{2\sigma^2}\right)
\end{equation}
where $\eta=\ln(N(H_2)/<N(H_2)>)$, $\sigma$ is the width of the PDF, and $\eta_0$ is its central position. The best fit, shown by the solid red curve on Figure~\ref{fig:pdf}, returns a value of $\sigma=0.34$. 

The higher $\eta$ part of the PDF can be fit with a power-law of the form $PDF \propto N(\mathrm{H}_2)^s$ with a best-fit index of $s=-2.48\pm0.06$ shown by the dashed red-line. If a region can be considered a single structure, then the power-law index $s$ is related to the form of a radial density profile $\rho\propto\eta^{-\alpha}$ by the relation $\alpha=1-2/s$ for a spherical distribution \citep{2013ApJ...763...51F}. The fitted $s$ thus implies a radial profile with an index of $\alpha=1.8$. Alternatively, for a cylindrical distribution, the relation is $\alpha=1-1/s$, which gives a profile index of $\alpha=1.4$. 

To examine the physical origins of the different parts of the PDF, we split the column density map into the region associated with HCL2 (the region contained with the dashed-box in Figure~\ref{fig:nh2}) and the material outside of that region (the area outside of the dashed-box). The HCL2 PDF is shown in Figure~\ref{fig:pdf} as the blue curve while the green curve shows the PDF of material excluding HCL2. The former  follows the previously fit power-law.
Meanwhile, the latter closely follows the previously fitted log-normal curve with only minimal deviation at the higher densities.

\subsection{Cloud Structure}
\label{sec:csar}
\begin{figure}
\includegraphics[width=\columnwidth]{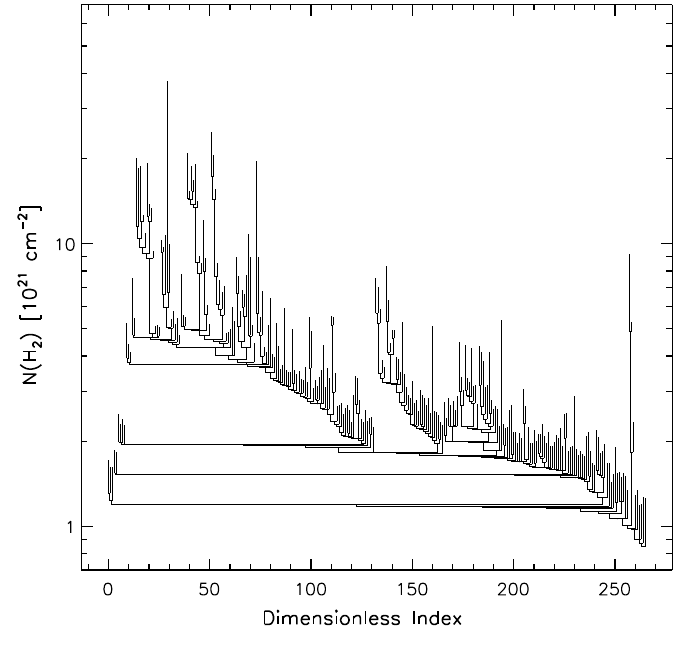} \\
\caption{\label{fig:dendro} A dendrogram of the \nh\ column density map towards the TMC1 region created with the CSAR algorithm \citep{2013MNRAS.432.1424K}. Horizonal lines are closed contours around resolved features with a minimum density contrast ($3\ \sigma$), vertical lines show the contrast between contours and their substructure. Upward lines without a capping horizontal are sources which have no resolved substructure. }
\end{figure}

In Section~\ref{sec:catalog}, we investigate the population of dense prestellar cores in the TMC1 region. These should appear in the maps as compact sources. Figure~\ref{fig:dendro} shows a tree (a dendrogram) of structure in the TMC1 region created from the \nh\ column density map using the CSAR source extraction algorithm \citep{2013MNRAS.432.1424K}. The dendrogram is a symbolic representation of the structure in the region. The horizontal lines are closed isocontours, and the vertical lines show the contrast between that contour and the contour level at which that region becomes substructured. Upward stems that are not terminated by a horizontal line are sources which do not contain resolved substructure.

The rms across visually flat, ${\sim}12\arcmin$-wide regions of the \nh\ map is $0.07-0.09\times10^{21}$~cm$^{-2}$. Therefore, we adopted for the decomposition shown in Figure~\ref{fig:dendro} a conservative $\sigma$ of $0.09\times10^{21}$~cm$^{-2}$, a minimum peak-to-edge contrast of 3 times this, and a minimum resolution (i.e., source size) equal to the HPBW of the telescope beam at 500 $\mu$m.

We find a total of 265 unstructured sources. These are enumerated by the dimensionless index (horizontal axis of Figure~\ref{fig:dendro}). The main structure in HCL2 appears as a network of branches above ${\sim}2\times10^{21}$\,cm$^{-2}$ between indices of $\sim10-130$. Examining the column density values in Figure~\ref{fig:pdf} and Figure~\ref{fig:mass}, we see that $\sim2\times10^{21}$\ cm$^{-2}$ is approximately the same column density where the PDF transitions from log-normal to linear and where the linearity of the cumulative histogram breaks down.

There are also hundreds of single-leaf branches that merge consecutively with the larger trees (e.g., the pattern around index 100 or 299--260). These branches correspond to minor cirrus and background features with little or no organised substructure.

\section{Filamentary Structure}
\label{sec:filaments}

\begin{figure*}
\includegraphics[width=\textwidth]{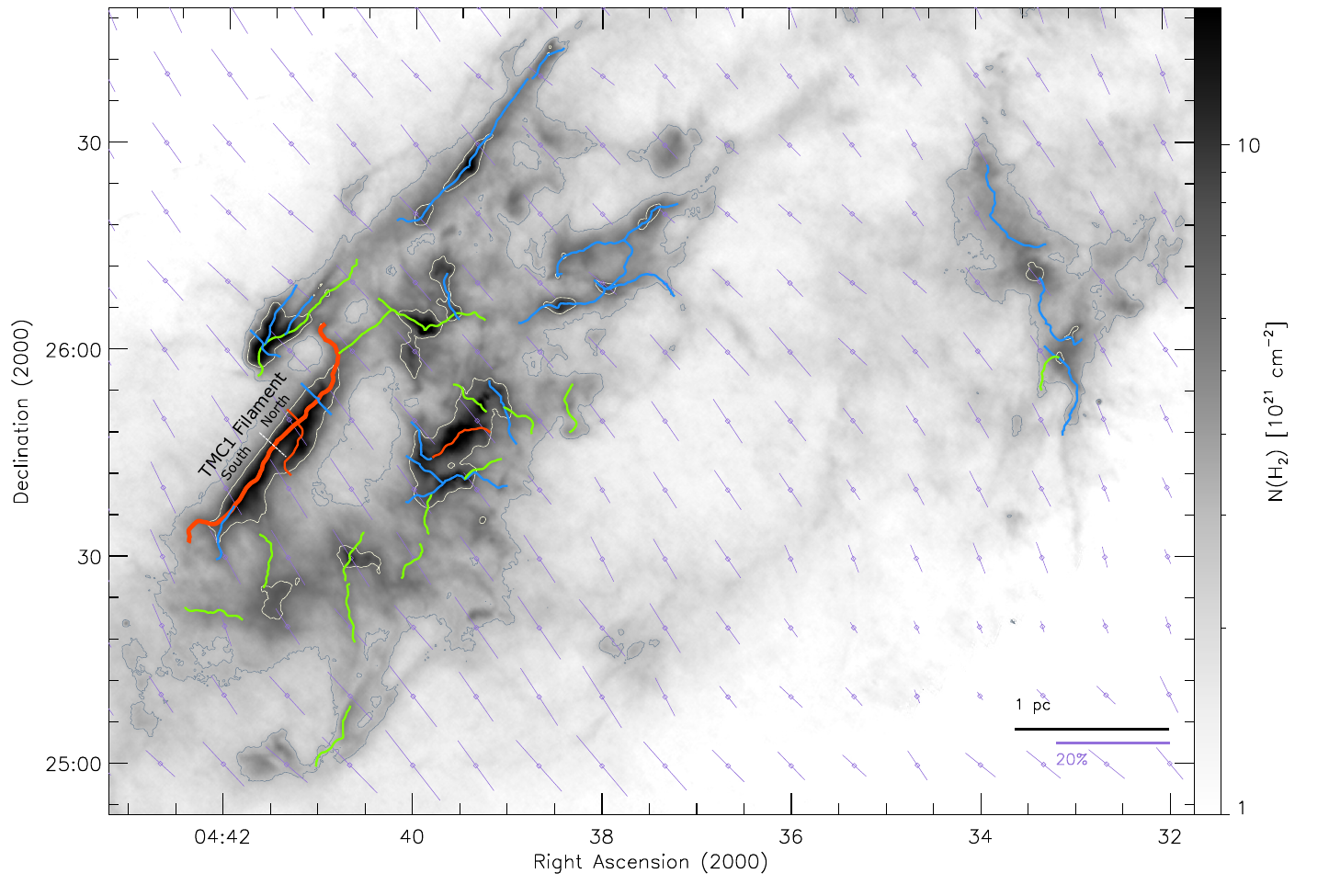}
\caption{\label{fig:fil} The filamentary network towards the TMC1 region as traced by \texttt{DisPerSE} plotted over the $N(\textrm{H}_2)$ map. The red, blue, and green lines denote filaments that are supercritical, trans-critical, and subcritical against axisymmetric perturbations, respectively (see text for details). The purple vectors show the magnetic field direction inferred from the 353 GHz {\it Planck} polarisation data. Distance and percentage polarisation reference bars are shown in the bottom-right corner. The grey and white contours denote column densities at the equivalent of an $A_\mathrm{V}$ of 3 and 7 mag, respectively. The TMC1 Filament is annotated and shown by a thicker line weight, and a short white orthogonal line shows the split between the North and South sections. 
}
\end{figure*}

\subsection{Extracting Filaments}
\label{sec:exfil}
One of the major features of the \textit{Herschel} era has been interest in filamentary modes of star formation  \citep[e.g.,][]{2014prpl.conf...27A,2019A&A...621A..42A,2020MNRAS.492.5420S}. There are several topological methods for tracing filamentary structures in maps. We have used here the \textsc{DisPerSE} software package to trace filamentary structures in the \nh\ map of the TMC1 region.

The \textsc{DisPerSE} package traces persistent structures by connecting critical points, e.g., maxima, saddle-points, and minima in a 2D map, with integral lines, e.g., tangential paths to the local gradient \citep{2011MNRAS.414..350S}. Key to this concept is persistence, i.e., the minimum allowed contrast between two connected critical points. \citet{2019A&A...621A..42A} examined the input parameters with respect to \textit{Herschel} maps and recommended a persistence value on the order of the background rms on the map.

Figure~\ref{fig:fil} shows the results for running \textsc{DisPerSE} on the low-resolution \nh\ map. We adopt a conservative persistence threshold of $0.09\times10^{21}$~cm$^{-2}$, equal to the rms used for the CSAR decomposition in Section~\ref{sec:csar}. We follow \citet{2019A&A...621A..42A} in using an assembly angle of $50^\circ$, a smoothing kernel of twice the map resolution and a minimum background contrast of $2\times10^{21}$~cm$^{-2}$, i.e., $1.5\times$ the mean background where the rms was measured. Features shorter than 10 times the map resolution (360\arcsec, $\sim0.25$~pc @ 140~pc) were removed from the network, leaving a total of 53 filaments.

Filaments were identified in the lower resolution (36\arcsec) column density map, but features were then measured from the higher-resolution (18\arcsec) map. For example, median radial profiles were constructed for each filament using the skeleton data from \textsc{DisPerSE}'s run. The orientation of the filament at each critical point was calculated by averaging the angle of the skeleton segments before and after it. A profile of the pixels orthogonal to this angle was extracted for each side the filament at the node. The median profile on each side of the filament was then taken, the dispersion at each point being the median deviation from the median profile.

Integrating across the filament profile gives $M_\mathrm{line}$, the mass per unit length of the filament. The critical $M_\mathrm{line}$ for an isothermal cylinder to be thermally stable against axisymmetric perturbations is
\begin{equation}
    M_\mathrm{line,crit}=\frac{2 c_\mathrm{s}^2}{G}=\frac{2 k_\mathrm{B} T}{G \mu m_\mathrm{H}},
\end{equation}
where $c_\mathrm{s}$ is the sound speed, $G$ is the Gravitational constant, $T$ is the gas temperature, $\mu$ is the mean mass per particle, and $m_\mathrm{H}$ is the mass of a hydrogen nucleus \citep{1997ApJ...480..681I}. For typical molecular weights ($\mu=2.37$, e.g., \citealt{2008A&A...487..993K}) and temperatures ($T=10$~K) in molecular star formation regions, the critical line mass is $16.2$~M$_\odot/pc$. We check the mean mass per unit length of each filament against this value to test each filament's stability. 

The 53 filaments were then filtered via the robustness criterion listed in \citet{2019A&A...621A..42A}, namely that the filaments have an axis ratio (length divided by Gaussian FWHM) greater than 3 and a background-to-filament amplitude contrast greater than 0.3. This final cut left 35 robust filaments out of 53. The robust filaments are shown in Figure~\ref{fig:fil}. These are colour coded to match the \citet{2019A&A...621A..42A} descriptions of thermally supercritical, transcritical, and thermally sub-critical against the critical mass to unit length. 
The single longest supercritical filament shown in Figure~\ref{fig:fil} is the TMC1 filament (it is annotated). This filament corresponds to 
an $\sim11$~K linear feature in Figure~\ref{fig:nh2}, demonstrating that it is one of the coldest and densest features in the TMC1 region. Figure~\ref{fig:fil} also illustrates that most of the other filaments in the TMC1 region are sub-critical filaments that generally appear at lower background column densities. 

\subsection{Filament Widths}
\label{sec:filwid}
\begin{figure}
\includegraphics[width=\columnwidth]{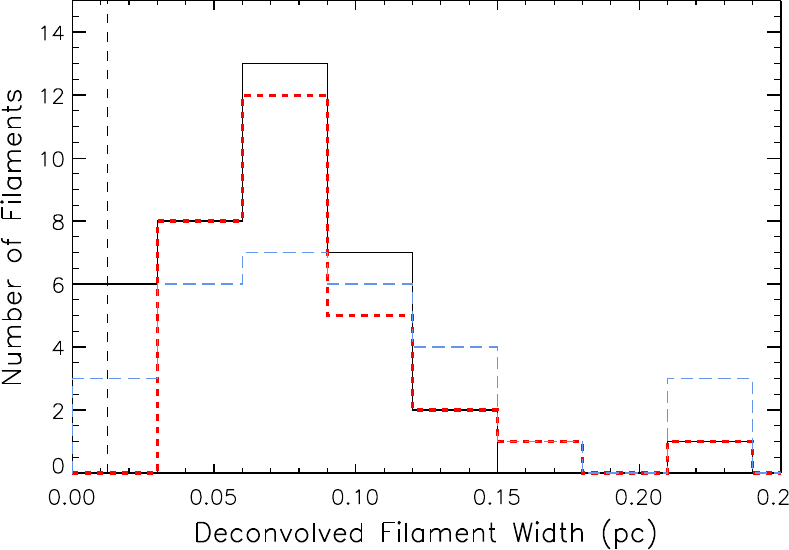} \\
\caption{\label{fig:FWHM} The distribution of deconvolved FWHM widths from Gaussian fitting to the inner part of each filament. The red-dotted and blue-dashed curves show the widths derived from the separate ``left'' and ``right'' fits for each filament. The black curve shows the FWHM averaged between left and right for each filament. The vertical black-dashed line shows the effective FWHM of the map. The FWHM were fitted on the high-resolution (18'') column density map. }
\end{figure}

The filament profiles exhibit a range of morphologies. The centres are broadly Gaussian, but a number of them depart from that shape in their outer parts and are better fit with a Plummer-like profile. To gauge the $FWHM$ of the inner part of the filament, we follow \citet{2019A&A...621A..42A} by fitting a Gaussian to the profile within $1.5\times\mathrm{HPHW}$ of the centre, where the $HPHW$ is the point where the filament profile has dropped to half the peak amplitude measured above the background. The effective beam $FWHM$ of the map is subtracted in quadrature from the fitted Gaussian to give a deconvolved filament width, $FWHM_\mathrm{dec}$. 

The histogram of measured $FWHM_\mathrm{dec}$ is shown in Figure~\ref{fig:FWHM}. The width of each filament is determined twice using each side of the median profile of the filament independently. The distribution of the ``left-hand'' and ``right-hand'' fits is shown in Figure~\ref{fig:FWHM}. The distribution of averaged $FWHM_\mathrm{dec}$ for all robust filaments is also shown. This latter histogram shows a peak at $\sim0.07$~pc, identical to the the peak in the filament width distribution found in the neighbouring B211/213 region \citep{2013A&A...550A..38P}. As discussed in \citet{2022A&A...667L...1A}, width estimates based on Gaussian fits are typically underestimated in the presence of significant power-law wings, and the deconvolution of widths at larger distances can make it hard to interpret filament FWHMs. However, the effective FWHM used for deconvolution (vertical line in Fig 9) is sufficiently small for these observations not to make an appreciable difference. Based on a Plummer-fit analysis, the TMC1 filament itself has a mean half-power width of $\sim0.13$~pc (see Section~\ref{tmc1filament} below).

\subsection{Filament Orientation}

\begin{figure}
\includegraphics[width=\columnwidth]{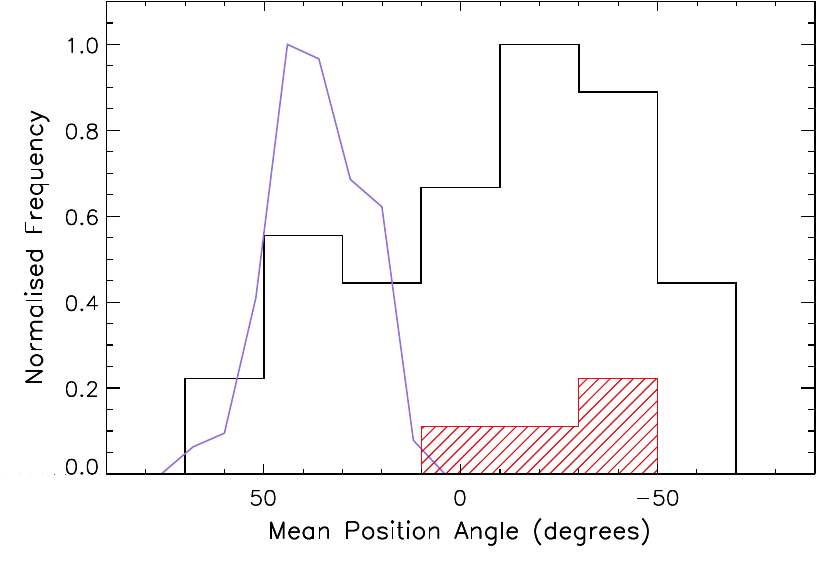} \\
\caption{\label{fig:orient} Comparison of filament position angles to the magnetic field direction. The black curve shows the histogram of all robust filaments towards the TMC1 region. The red-shaded histogram shows the orientation of supercritical filaments. The purple curve shows the normalised histogram of the magnetic field direction across the map sampled on a 1\arcmin\ scale. Both red and black histograms are normalised to the peak of the black histogram.}
\end{figure}

We compare the orientation of the TMC1 regions filaments with the larger-scale magnetic field direction inferred from the Planck 353 GHz polarisation products \citep{2020A&A...641A..11P}. The purple vectors in Figure~\ref{fig:fil} show the directions of the larger-scale magnetic field so determined. The field is broadly uniform across the TMC1 region but does bend slightly. The majority of supercritical filaments appear to be orthogonal to the field direction. 
    
We define the orientation of a filament as the median position angle across its nodes. The normalised distribution of mean position angles is shown in Figure~\ref{fig:orient}. The relative distribution for supercritical filaments occupies a narrow position angle range with values of approximately $-50-0^\circ$. The normalized orientation of the magnetic field vectors from Figure~\ref{fig:fil} is also shown. There is a clear offset between these two position angles of approximately $90^\circ$ degrees, showing that the supercritical filaments are orientated orthogonal to the magnetic field direction. This result is similar to that of \citet{2013A&A...550A..38P} who found that the main supercritical filament in the L1495 region was also aligned orthogonal to its local magnetic field direction. 

\subsection{The TMC1 Filament}
\label{tmc1filament}
\begin{figure*}
\begin{center}
    \includegraphics[width=0.9\textwidth]{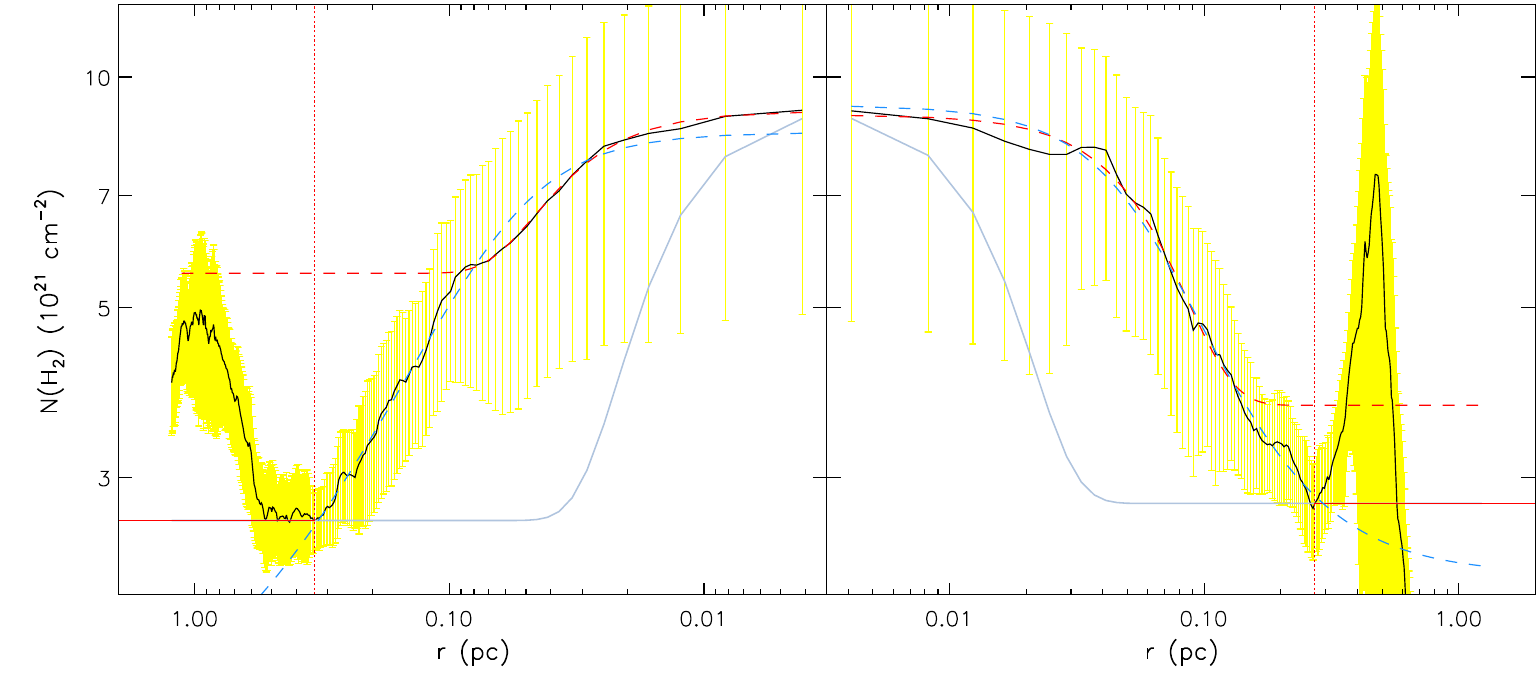} \\
    \includegraphics[width=0.9\textwidth]{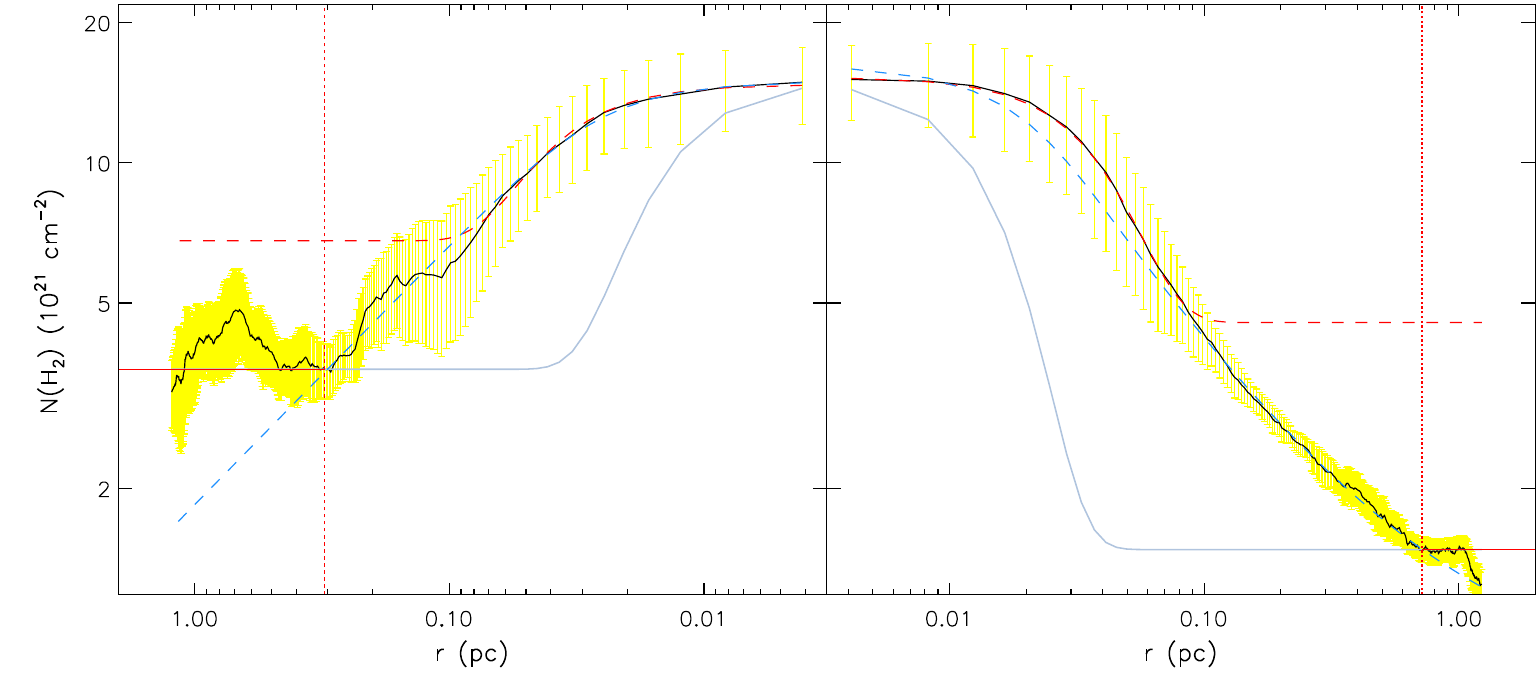}
\end{center}
\caption{\label{fig:fill1} Median left (NE) and right (SW) column density profiles of the North and South parts of the TMC1 filament as measured from the high-resolution (18'') column density map. Properties and fits are done independently for each side of each filament. The black curve is the median column density profile itself and the yellow bars are the median dispersion along the length of the profile at that radius. The red-dashed curve is a Gaussian fit to the inner profile. The horizontal solid red line is the background column density level (where the smoothed gradient of the profile is approximately zero) and the vertical dotted-line is the filament width at this background level. The dashed blue line is a best fit Plummer-like profile. The thin blue line is the resolution of the column density map. See text for full details.  }
\end{figure*}

The TMC1 filament is a thermally supercritical filament that runs in a north-westerly direction from an R.A. of 04:42 and Dec. of +25:35 to an RA of 04:41 and Dec. of +26:00, as shown and annotated in Figure~\ref{fig:fil}. This filament is heavily studied as a prototypical chemically young star-forming site \citep{2018Sci...359..202M,2018ApJ...854..116S,2019A&A...624A.105F}. The \textsc{DisPerSE} extraction split the TMC1 filament in two at a Declination of +25:47:45. We refer to the two parts as the ``North'' and ``South'' TMC1 filament, each of which has a length of 0.9~pc. Note that the North and South filaments respectively correspond to the distinct Planck Galactic Cold Clumps G174.20-13.44 and G174.40-13.45 \citep{2016A&A...590A..75F,2016A&A...594A..28P}. 

As noted earlier, \citet{2008MNRAS.384..755N} found that the TMC1 filament (the ``Bull's Tail'') was consistent with a narrow, cold ($\sim8$~K) core and a broader, warmer ($\sim12$~K) jacket. The \textit{Herschel} emission is a product of dust at different temperatures along the line of sight so the single temperature estimated from it will be a characteristic average. Nevertheless, the Bull's Tail is prominently shown as a cold feature (the red strip in Figure~\ref{fig:nh2}). The filament is sufficiently dense to have been seen in absorption in a \textit{Spitzer}~70 $\mu$m map \citep{2008MNRAS.384..755N}, as discussed in Section~\ref{coldens}.

The median profiles for the North and South filaments are shown in Figure~\ref{fig:fill1}. The profiles have been reconstructed and analysed independently on each side of each filament, following the approach and conventions of \citet{2019A&A...621A..42A}. Both filaments run in a broadly SE to NW orientation. Therefore, the left hand profiles in Figure~\ref{fig:fill1} correspond to the NE side of the filaments, whereas the right hand profiles correspond to the SW side. For simplicity, we refer to these as the ``left'' and ``right'' profiles.

Figure~\ref{fig:fill1} shows the median profile along the filament and the median dispersion around the median profile. In addition, the best fit Gaussian profile to the inner part of the profile is shown (see Section~\ref{sec:filwid}). The South filament (both sides) and the left North filament have $FWHM_\mathrm{dec}$ approximately equal to the peak of 0.07~pc shown in Figure~\ref{fig:FWHM}, but the right North filament has a larger $FWHM_\mathrm{dec}$ of 0.12~pc. The North filament is relatively well fit by a Gaussian profile out to large radii, albeit at a cost of poor constraint on peak. By contrast, the South filament is well fit on peak, but its greater peak-to-background contrast more readily shows a departure from the Gaussian profile. 

An estimate of the outer width of each profile can be made as the place where the gradient of the smoothed profile goes to zero (see \citealt{2019A&A...621A..42A} for details). This point is shown in each case also in Figure~\ref{fig:fill1}. The level of the profile at this point it taken as the profile background. The North filament appears broadly symmetric with similar background levels and widths on either side. The South filament, however, appears more asymmetric with a high background on the left and a low background on the right as it tapers away into the noise. The mean background level of the first three is $\sim3\times10^{21}\mathrm{~cm}^{-2}$, but is only half that for the last one. This difference can be seen in Figure~\ref{fig:fil} by the empty space between the filament and the edge of the figure. The North filament and the left profile of the South filament have widths of $\sim0.3$~pc. The right profile of the North filament, however, has a width of twice this at 0.7~pc, again, most likely due to the lack of other features surrounding it. 

The North filament has a mean column density  of $6.5\times10^{21}\mathrm{~cm}^{-2}$ whereas the South one has a mean of $13\times10^{21}\mathrm{~cm}^{-2}$. This difference is consistent with maps of molecular line emission that show stronger integrated emission from the South filament \citep[e.g.,][]{2016A&A...590A..75F,2018ApJ...864...82D}.

A Plummer-like profile of the form:
\begin{equation}
    N_{\textrm{H}_2}(r) = N_{\textrm{H}_2}(0)\left( 1+(r/R_\textrm{flat})^2 \right)^{-\frac{p-1}{2}} 
\end{equation}
was fitted to the background-subtracted profiles where $N_{\textrm{H}_2}(0)$ is the peak of the profile, $R_\mathrm{flat}$ is the approximate radius of the central region of constant column density, and $p$ is the power-law exponent of the column density profile at large values of $r$. Figure~\ref{fig:fill1} shows the best Plummer-like profile fit to each profile. The power-law exponents for all four profiles are in the range 1.6--2.6. Notably, the most ``featureless'' profile (lower right) has an exponent closest to 2 ($p=1.9$). The North and South part of the TMC1 filament have mean half-power widths from the Plummer fits ($~2\sqrt{3} R_\textrm{flat}$) of 0.20 and 0.06 pc respectively, the mean of these is $\sim0.13$\,pc. This does agree with the visual inspection of Figures~\ref{fig:fil} and ~\ref{fig:fill1} as there does appear to be more material around the Northern filament. 

The North filament has a mean $M_\mathrm{line}$ of 38~M$_\odot$/pc and the South filament a mean $M_\mathrm{line}$ of 69~M$_\odot$/pc. These values make them respectively 3 and 5 times thermally supercritical. These values are slightly higher than, but still consistent with, those found by \citet{2016A&A...590A..75F}. Figure~\ref{fig:cores} shows the location of dense cores in TMC1 (see next section). Interestingly, the North filament is associated with 6-7 dense cores, whereas the more thermally supercritical South filament is only associated with 1-2 dense cores. In both cases, only one core in each set is prestellar in nature (see the following Section for methodology). Zeeman measurements towards the South filament suggest that it is magnetically marginally supercritical as well, with a mass-to-flux ratio of 2.2 \citep{2019PASJ...71..117N}. 

Different velocity components have been detected within the entire TMC1 filament \citep{2016A&A...590A..75F,2018ApJ...864...82D}. These components have been interpreted as sub-filaments whose mass-per-unit length still exceeds thermal supercriticality  \citep{2016A&A...590A..75F,2018ApJ...864...82D}. The TMC1 filament therefore appears to be poised for collapse, with the North filament already having formed several dense condensations but the South filament still relatively undifferentiated with only a couple of dense condensations. This behaviour is consistent with the idea of TMC1 being a chemically young star-formation site \citep[e.g.]{2018Sci...359..202M,2018ApJ...854..116S,2019A&A...624A.105F}.

\section{Dense Core Catalogue}
\label{sec:catalog}

\subsection{Catalogue Creation}

The dendrogram in Figure~\ref{fig:dendro} and the filament maps in Figure~\ref{fig:fil} show that the dense cores in the TMC1 region are part of a complex tree of emission from which they need to be separated. For the creation of a dense core catalogue, we use the \textsc{getsources} source extraction package which can separate blended sources and work simultaneously across multiple wavebands \citep{2012A&A...542A..81M}.

In brief, \textsc{getsources} spatially decomposes the emission in a map into a series of strongly filtered ``single-scale'' images. A source is then identified on the smallest spatial scale and tracked through larger spatial scales until it reaches some maximum specified size, or its surface brightness becomes too low to be detected \citep{2012A&A...542A..81M}. The software is run in two passes -- the first for detection, the second for measurement. Decomposing a set of observations in this manner allows equivalent spatial scales from multiple wavebands to be correlated together during the actual extraction process and thus eliminates the need to cross-match independent monochromatic catalogues after the fact \citep{2012A&A...542A..81M}.

The \textit{Herschel} Gould Belt Survey was primarily designed to take a census of cold starless cores which primarily emit at longer wavelengths (160 -- 500$\mu$m) and embedded YSOs and protostellar cores which also emit at shorter wavelengths (e.g., 70 $\mu$m). It was therefore necessary to create two source extraction forks optimised for these two classes of object. Version v1.140127 of \textsc{getsources} was used. The \textit{Herschel} Gould Belt Survey catalogue creation procedure is described in full in \citet{2015A&A...584A..91K}.

A common extraction was performed up to the detection stage on flattened backgrounds. The first extraction fork created an extended source catalogue by using the 160-500$-\mu$m maps for source selection. The second extraction fork created a 70-$\mu$m-identified compact source catalogue using only the 70-$\mu$m map for selection. For both catalogues, photometry was performed at 70-500~$\mu$m. The extractions' focus on the dense material was enhanced by using the high-resolution column density map as a pseudo-wavelength, and a temperature-corrected 160-$\mu$m map. These weighted the extraction towards sources of high column density rather than just high surface brightness.

Fifty-five sources were extracted in the compact source pass, and 510 were extracted in the extended source pass. Of those, 45 and 285 were defined as acceptable, respectively, based on their global detection significance \citep[\texttt{GOOD=1}, see][for details]{2012A&A...542A..81M}. \textsc{getsources} separates the detection significance of extraction from the photometric significance of the resulting measurements. These are respectively SIG\_MONO (the best signal-to-noise on a single spatial scale) and SNR\_MONO (the traditional photometric signal-to-noise).

A set of selection rules was applied to each of the photometric catalogues to create science catalogues. A catalogue of dense core candidates was created by filtering the extended source catalogue for sources that:
\begin{enumerate}
\item were detected (SIG\_MONO$>5$) in at least two wavelengths and the $N(H_2)$ map,
\item had reliable flux measurements (SNR\_MONO$>1$) in at least one wavelength and the $N(H_2)$ map,
\item and a detection significance summed over all wavelengths of at least 10.
\end{enumerate}
These filters produced a catalogue of 201 objects. These sources were visually inspected, and 23 entries (11\%) were rejected, leaving a final catalogue of 178 dense core candidates.

The robustness of these dense core candidates was assessed by comparing the \textsc{getsources} extraction to the results of a \textsc{CSAR} extraction performed on the high-resolution column density map. We find that 48\% of sources in the extended source catalogue matched a CSAR position within their $N(H_2)$ FWHM contour, and a further 33\% of starless core candidates matched a CSAR core. These percentages are typical of those found in other clouds. 

Starless cores are pockets of molecular gas that do not have embedded protostellar sources. The dense core catalogue was filtered for these by comparing it against the compact 70-$\mu$m source catalogue from the compact extraction, the \textit{Spitzer} Taurus YSO catalogues of \citet{2010ApJS..186..259R}, and the NED Extragalactic database \citep{1991ASSL..171...89H}. In all cases, a match was found if external  catalogue coordinates were found to lie within the half-power ellipse of a dense core in the high-resolution \nh\ map. A total of eleven sources, ten YSO candidates, and one galaxy, were rejected by this process, leaving a final catalogue of 167 starless core candidates.

\subsection{Core Properties}

The properties of TMC1 region starless cores are listed in Table~\ref{tab:prop} and their locations are shown in Figure~\ref{fig:cores}

We define two radii for the cores. The raw radius of the core is the geometrical average of the major and minor FWHMs of the core in the high-resolution column density map, $\overline{FWHM}$. The deconvolved radius is then calculated as $R_\textrm{decon}=\sqrt{\overline{FWHM}^2 - \overline{HPBW}^2}$ where $\overline{HPBW}$ is the resolution of the high-resolution column density map (18.2\arcsec).

The temperatures and masses of the starless cores were estimated by fitting their integrated flux SEDs between 160 and 500 $\mu$m in the same manner used for the creation of the column density map (see Section~\ref{coldens} for details). Following \citet{2015A&A...584A..91K}, the quality of the fit was checked by fitting a second permutation of the SED using the wavelengths 70 $\mu$m to 500 $\mu$m and by weighting via the detection significance rather than the measurement errors. SED results were accepted if the two mass estimates were within a factor of two of each other, at least three bands were used to fit the SED, and the integrated flux at 350 $\mu$m was higher than that at 500 $\mu$m. This process left 59\% of the cores with a valid SED, i.e., 96 cores. A temperature equal to the median temperature of cores with valid SEDs is assumed for cores without a valid SED (see Section~\ref{sec:temp}).

\subsection{Prestellar Cores}

\begin{figure*}
\includegraphics[width=\textwidth]{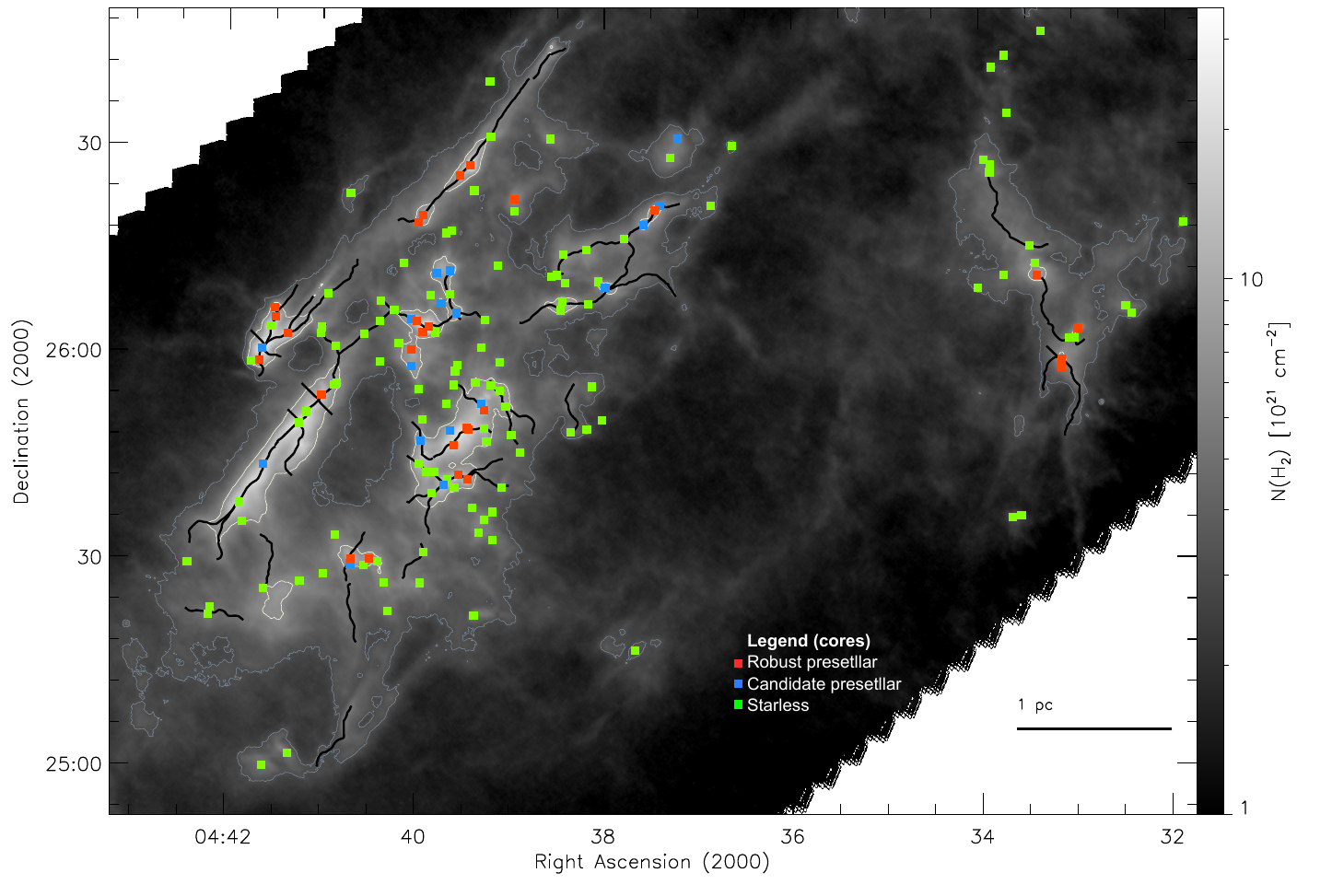} \\
\caption{\label{fig:cores} The location of starless and prestellar cores overplotted on the \nh\ map. The red, blue, and green markers show the respective locations of robust prestellar, candidate prestellar, and starless cores. Two contour levels, $A_V = 3$ and $7$,  are shown for comparison. The network of filaments from Section~\ref{sec:filaments} is also shown for comparison.}
\end{figure*}

\begin{figure}
\includegraphics[width=\columnwidth]{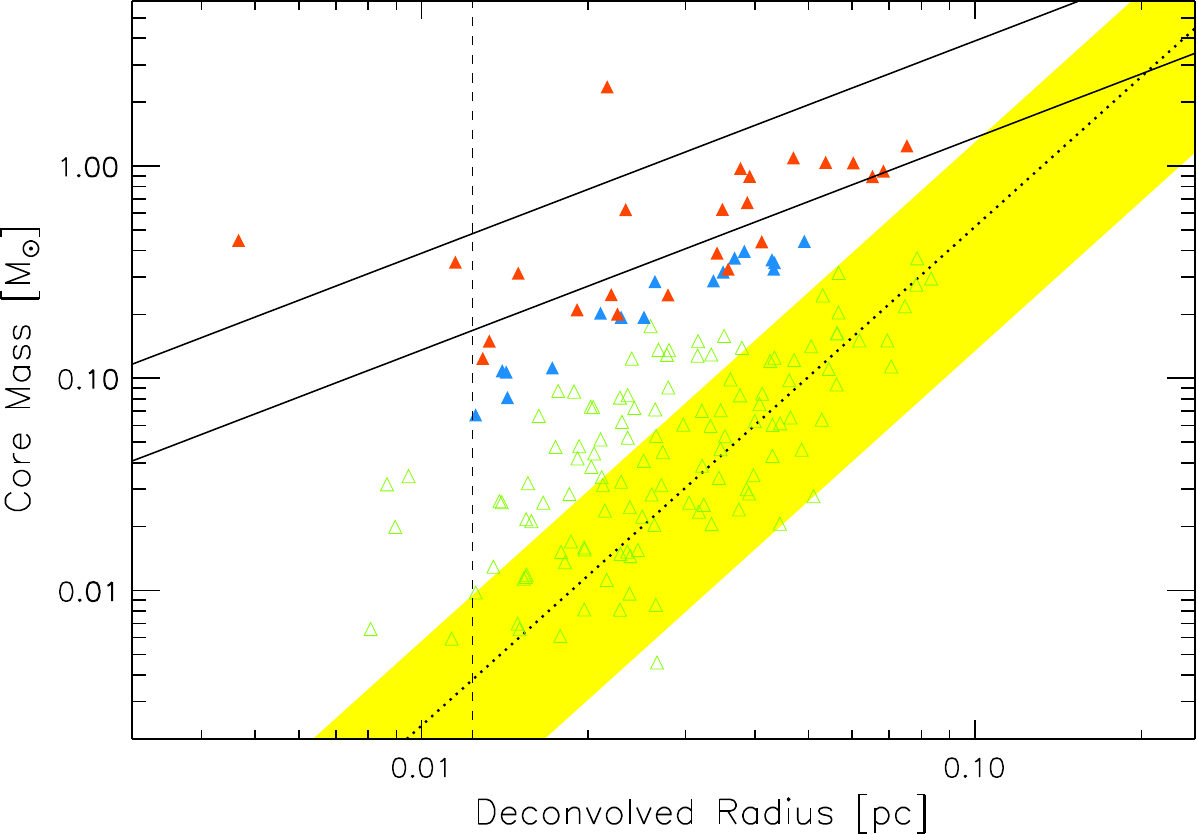} \\
\caption{\label{fig:masssize} Core mass versus size for the population of starless cores towards the TMC1 region. The green symbols show unbound starless cores, the blue symbols candidate prestellar cores, and the red symbols show robust prestellar cores. The size is taken as the deconvolved radius from the high-resolution column density map while the mass is the mass from greybody fit to the integrated fluxes of the cores. The yellow band and the dotted line shows the expected mass-size relationship for diffuse CO clumps \citep{1996ApJ...471..816E}. The solid lines show the expected mass-size relationship for 7-K and 20-K BE spheres. }
\end{figure}

Starless cores which are likely to be taking part in the star formation process are termed prestellar (i.e., they have some signs of being gravitationally bound). An estimate of the stabilities of the cores can be made by comparing their masses to that of a critical Bonner-Ebert (BE) sphere \citep{1955ZA.....37..217E,1956MNRAS.116..351B}. A BE sphere is a pressure-confined isothermal sphere of gas that will be unstable against collapse if its mass is above a critical value given by,
\begin{equation}
M_\mathrm{crit} = 2.4 \frac{R c_\mathrm{s}^2}{G}
\end{equation}
when the sound speed is given by
\begin{equation}
c_\mathrm{s}^2 = \frac{ k_\mathrm{B} T}{m_\mathrm{H} \mu}
\end{equation}
and $R$ is the sphere's radius, and the other values are as defined earlier. We estimate $M_\mathrm{crit}$ for each core using its deconvolved radius for $R$ and its catalogue temperature for $T$.

Following \citet{2015A&A...584A..91K} and with analogy to the virial ratio, we classify a starless dense core as a robust prestellar core if the ratio $\alpha=M_\textrm{crit}/M_\textrm{core}$ is $\leq2$. In practice, marginally resolved cores will have a significant error in $R$. Therefore, following \citet{2015A&A...584A..91K} and \citet{2016MNRAS.459..342M}, we adopt a size-dependent threshold given by $\alpha\leq 5\times (\overline{HPBW}/\overline{FWHM})^{0.4}$ where $\overline{HPBW}$ is the map resolution and $\overline{FWHM}$ is the mean undeconvolved source radius (see above) measured in the high-resolution \nh\ map. Cores that meet this second criterion but not the first are classified as candidate prestellar cores. 

Out of the catalogue of 167 starless cores, 27 were classified as robust prestellar cores, and a further 17 were classified as candidate prestellar cores. Figure~\ref{fig:cores} shows the location of the starless, candidate prestellar, and robust prestellar cores. As shown by the extinction contours, most of the robust and candidate prestellar cores appear at locations with a column density higher than $A_V=5-7$ and close to the location of filaments. 

Figure~\ref{fig:masssize} shows a plot of deconvolved core radius versus mass for all cores towards the TMC1 region. The unbound starless cores cluster towards the lower part of the diagram and are coincident with the yellow band which shows the mass-size correlation observed for diffuse CO clumps \citep{1996ApJ...471..816E}. The robust prestellar cores and candidate prestellar cores, however, have relatively higher masses.

\subsection{Temperature and Density of Cores}
\label{sec:temp}

\begin{figure*}
\includegraphics[width=0.49\textwidth]{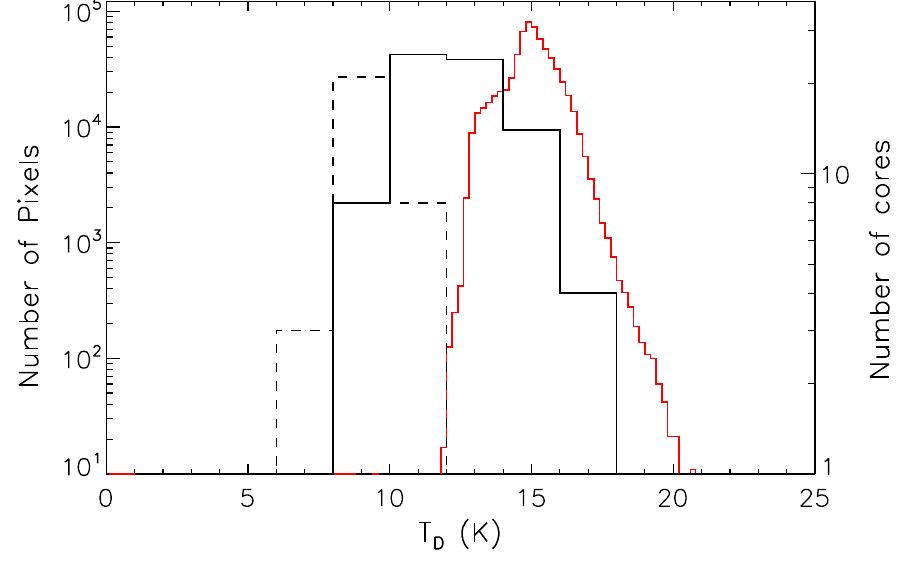}
\includegraphics[width=0.49\textwidth]{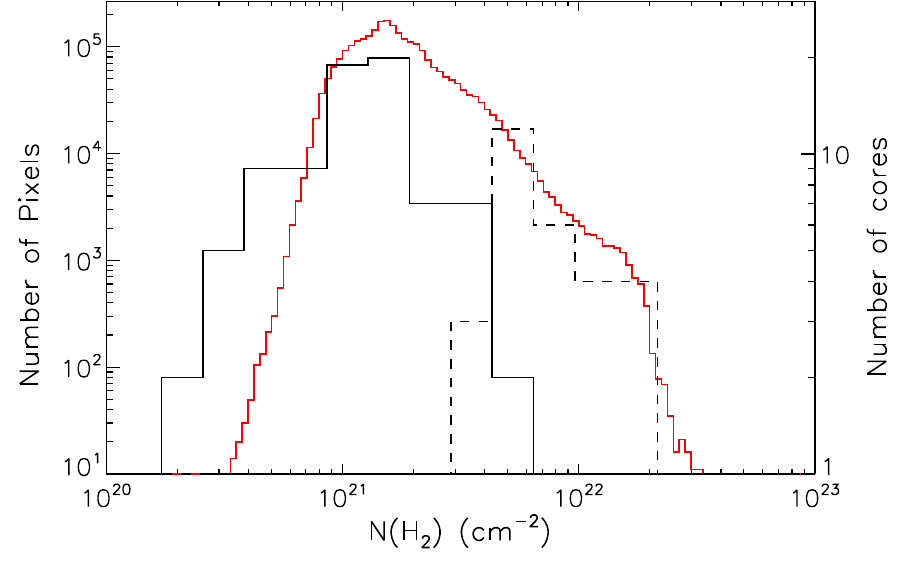}
\caption{\label{fig:sedprop} Histograms of \textbf{(left)} fitted SED temperatures and \textbf{(right)}  core background-subtracted peak \nh\ column densities at the 500$\mu$m resolution. The right-hand axis and the black curve on each plot show the histogram for unbound starless cores (solid line) and prestellar cores (dashed line). The left-hand axis and the red curve on each plot show the histogram of all valid pixel values in the map from Figure~\ref{fig:nh2}.  }
\end{figure*}

\begin{figure}
\begin{center}
\includegraphics[width=0.8\columnwidth]{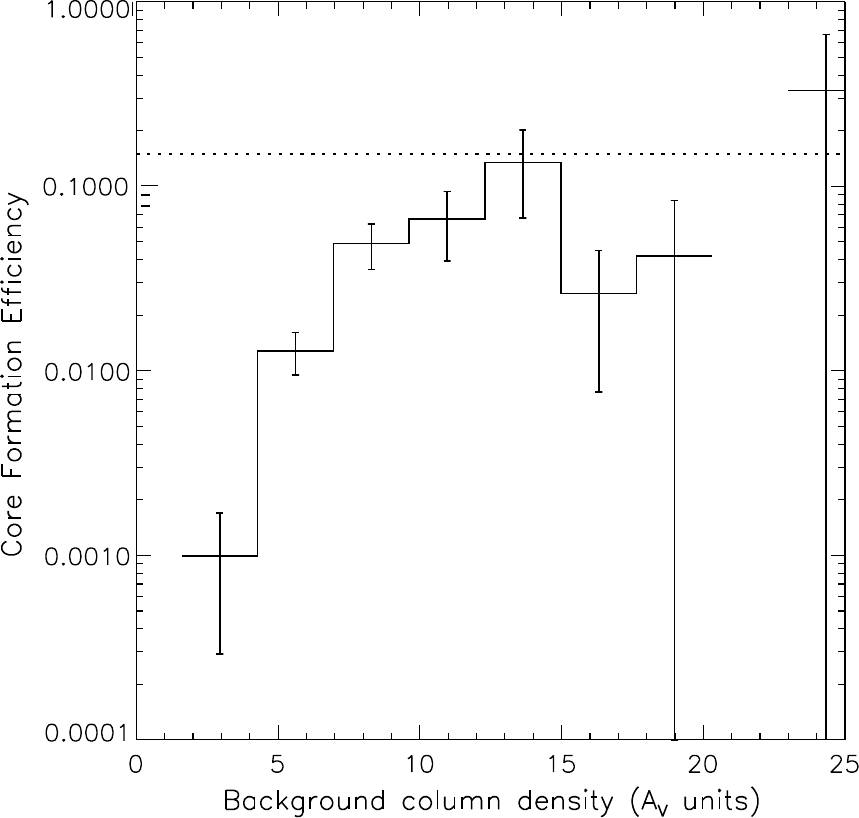}    
\end{center}
\caption{\label{fig:bgeff} Core formation efficiency versus extinction. The horizontal dashed line is an efficiency of 15\%. }
\end{figure}

The left-hand side of Figure~\ref{fig:sedprop} shows the distribution of temperatures from (valid) core SEDs for starless cores and prestellar cores. The median value of their combined distribution is $11.4\pm1.4$~K. This temperature is virtually identical to that found by \citet{2015A&A...584A..91K} for Aquila cores when they first used this technique. For cores without a valid SED fit, we estimate a mass from their longest significant wavelength assuming a dust temperature equal to this median. Figure~\ref{fig:sedprop} also shows the distribution of pixel-by-pixel temperatures in the map of SED dust temperatures (i.e., Figure~\ref{fig:nh2}). The median value of this distribution is $15.0\pm0.8$~K. These temperatures are generally higher than those for individual cores as the warmer inter-core medium dominates the latter distribution. The background flux around each core, including the contribution from the warmer dust, is subtracted from the cores by {\sc getsources} before SED fitting.

For each core, we also fit its background-subtracted peak intensity at a common HPBW (equivalent to that at 500 $\mu$m) to estimate the peak column density in the core. These peak column density estimates are plotted on the right-hand panel of Figure~\ref{fig:sedprop}. Shown here are the pixel values from the 500 $\mu$m resolution column density map and the fitted peak values from the cores. This panel shows that prestellar and unbound starless cores have distinctly different peak densities with a break at approximately $3-4\times10^{21}$~cm$^{-2}$ ($A_V \sim 3$\ mag). This column density is approximately the same as the break seen in the cumulative core mass function and the transition between the log-normal and power-law components of the PDF towards the TMC1 region. 

The differential core formation efficiency (CFE) can be used to examine the density at which a cores actually forms. The CFE is the ratio of the mass of cores forming at a specific background column density to the total mass in the cloud at that column density. In the regions of Aquila, Ophiuchus, and Cepheus, the CFE peaks at around 10--15\% between $A_V\approx10-20$ and levels off at higher extinctions \citep{2015A&A...584A..91K,2020ApJ...904..172D,2020A&A...638A..74L}. The dependence of the CFE on background column density is very similar in all those regions. What separates them is the low-$A_V$ edge at which they first hit 10--15\%. In Cepheus \citep{2020ApJ...904..172D}, Ophiuchus \citep{2020A&A...638A..74L}, and Aquila \citep{2015A&A...584A..91K}, this is $A_V = 7, 10$, and $20$, respectively. Figure~\ref{fig:bgeff} displays the CFE plot for the Taurus region, showing an initial jump at $A_V\sim5$ before reaching the 15\% level around $A_V\sim14$. Thus, the TMC1 region follows the same basic trend as the other regions, but there 
is not enough high column density material in Taurus/TMC1 to give reasonable statistics at higher extinctions. Figure~\ref{fig:sedprop} also reinforces the lack of high column density gas, and therefore the lack of cores at higher extinctions in TMC1, where the 
maximum column density in the region is $2\times10^{22}$\,cm$^{-2}$ ($A_V\sim22$). 

In the TMC1 region, star formation happens like in other parts of Taurus, such as L1495, but at somewhat lower extinctions compared to other clouds such as Aquila or Ophiuchus. There is only one thermally supercritical filament and only a few transcritical filaments in the region (as seen in Figure~\ref{fig:fil}). Only 10\% of all filaments have an average equivalent $A_v$ above $7-8$ mag, which is considered a density threshold for star formation. \cite{2020ApJ...904..172D} discuss the issue of core formation at low $A_V$ with respect to the Cepheus region and note that the relative unavailability of transcritical and supercritical filaments at lower column densities is the simplest explanation for the low number of cores.  It is possible that the same bottleneck applies in the TMC1 region, given its range of column densities. Even though there are fewer supercritical/transcrital filaments, a comparison of Figure~\ref{fig:fil} and Figure~\ref{fig:cores} does suggest that most robust/candidate prestellar cores of TMC1 lie along those filaments, in agreement with the filament paradigm of star formation \citep[cf.][]{2014prpl.conf...27A}.

\subsection{Completeness for Prestellar Cores} 

\begin{figure}
\includegraphics[width=\columnwidth]{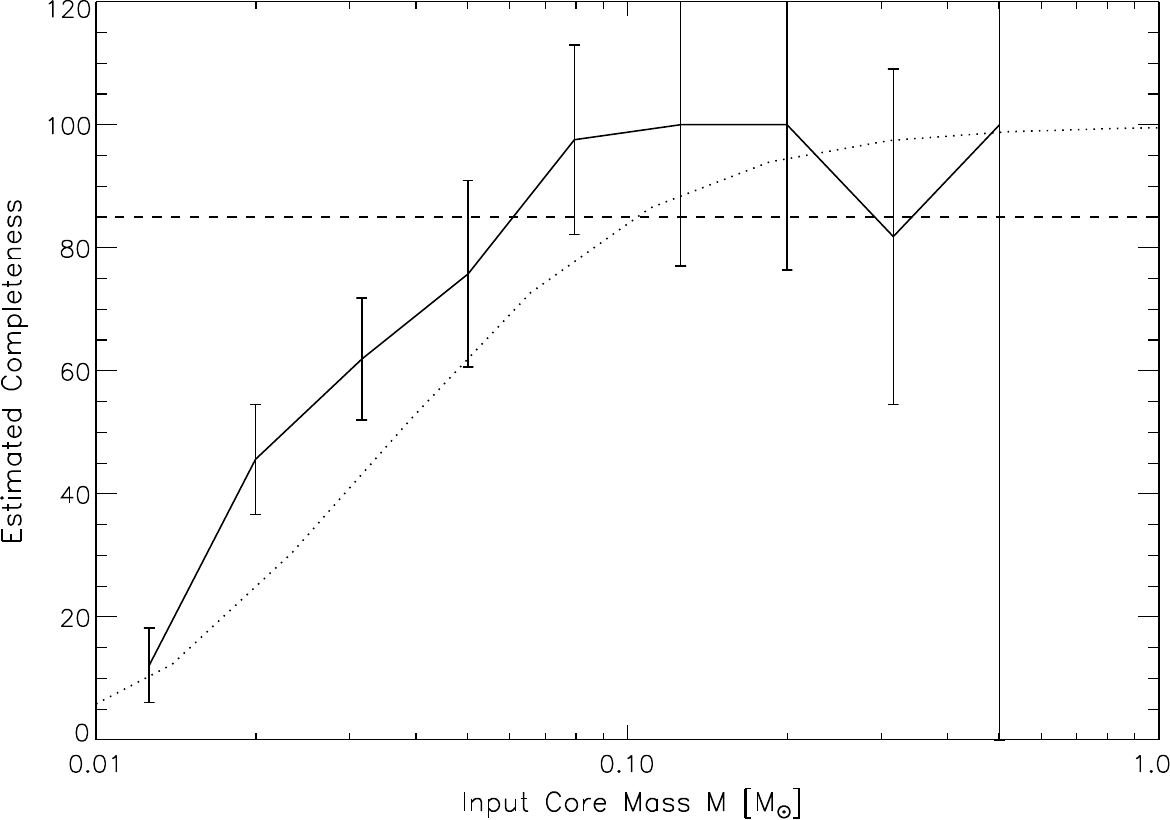}
\caption{\label{fig:comp} The fraction of recovered simulated prestellar cores versus simulated core mass. The error bars assume a Poisson distribution. The horizontal dashed-line shows 85\% completeness. The dotted-line shows the completeness model from \citetalias{2016MNRAS.459..342M} and Appendix B.2 of \citet{2015A&A...584A..91K}. }
\end{figure}

We estimate the completeness of our prestellar core extraction by performing Monte Carlo simulations following \citet{2015A&A...584A..91K}. The details and simulated core population are the same as used in \citetalias{2016MNRAS.459..342M}. In brief, our detected sources were removed from the \textit{Herschel} flux maps and a population of 267 simulated critical BE spheres were injected into that map towards regions of high column density. A full \textsc{getsources} extraction was then performed on the simulated maps. The completeness tests were focused on the range of masses where the completeness limit was expected to reside based on other \textit{Herschel} studies \citep{2015A&A...584A..91K,2016MNRAS.459..342M,2018arXiv180107805B}. 

Figure~\ref{fig:comp} shows a plot of simulated core mass versus estimated completeness. A simulated prestellar core is considered detected if its detected position matches the input position within 9\arcsec\ and if it is classified as a prestellar core via the process described earlier in this Section. A total of 181 out of the 267 simulated cores were recovered and correctly classified as prestellar by this procedure.  The horizontal dashed line in Figure~\ref{fig:comp} shows the same 85\% completeness threshold as used in \citetalias{2016MNRAS.459..342M}. The estimated completeness drops below this threshold consistently for simulated cores with masses $<0.06$~M$_\odot$. The dotted-line in Figure~\ref{fig:comp} shows the same model for completeness as used in  \citetalias{2016MNRAS.459..342M} based on the procedure described in \citet{2015A&A...584A..91K} and Appendix B.2 of \citet{2015A&A...584A..91K}.

Comparing the simulated to derived masses shows that the cores' masses are underestimated by approximately 10-20\% above this completeness level, albeit with considerable scatter. The standard deviation in the ratio of recovered to simulated masses settles down to 25--30\% for detected masses above 0.15~M$_\odot$. The value of $0.06$~M$_\odot$ from the previous paragraph could show that our extraction went marginally deeper than that in \citetalias{2016MNRAS.459..342M} which arrived at a completeness estimate of $M>0.1$~M$_\odot$ for detection. The difference could be ascribed to the slightly different dynamic range of the two tiles as that is an important factor in the sensitivity of a \textsc{getsources} extraction. Nevertheless, we adopt a conservative completeness threshold for detection of $M>0.1$~M$_\odot$ for consistency with \citetalias{2016MNRAS.459..342M} and a threshold for reliable mass determination above $M>0.15$~M$_\odot$. 

\subsection{Prestellar Core Mass Function}

\begin{figure}
\includegraphics[width=\columnwidth]{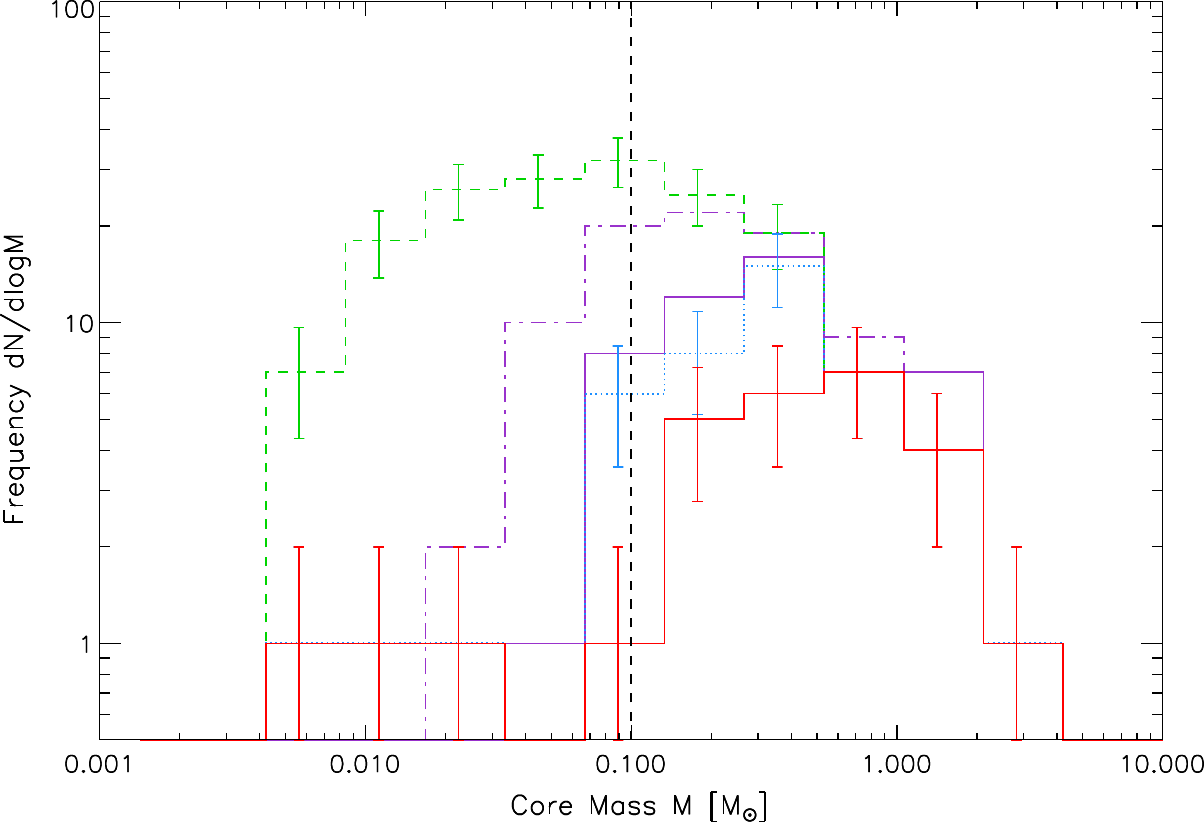} 
\caption{\label{fig:cmf} Histogram of starless core masses in the TMC1 region. The coloured lines denote different populations: dashed-green shows all starless cores, blue-dotted shows candidate and robust prestellar cores, while solid-red robust prestellar cores alone. The solid and dashed purple curves respectively show the robust and candidate cores from \citetalias{2016MNRAS.459..342M}. Error bars are shown based $\sqrt{N}$ uncertainties. The vertical dashed-black line shows our $0.1$~M$_\odot$ completeness limit.}
\end{figure}

Figure~\ref{fig:cmf} shows the core mass function (CMF) for our 44 prestellar cores (robust and candidate) and just the 27 robust prestellar cores. The median mass of the prestellar cores is 0.33~M$_\odot$. Figure~\ref{fig:cmf} also shows the CMF for all 167 starless cores in the TMC1 region. The median mass of the starless cores is 0.07~M$_\odot$, a mass just a little lower than the $0.1$~M$_\odot$ completeness limit. For comparison, we also shown in Figure~\ref{fig:cmf} the CMF for cores from the neighbouring L1495 region, as extracted in \citetalias{2016MNRAS.459..342M}. Note that L1495 contains more prestellar cores overall, largely due to the B211/213 filament and the cores associated with it.

\begin{figure}
\includegraphics[width=\columnwidth]{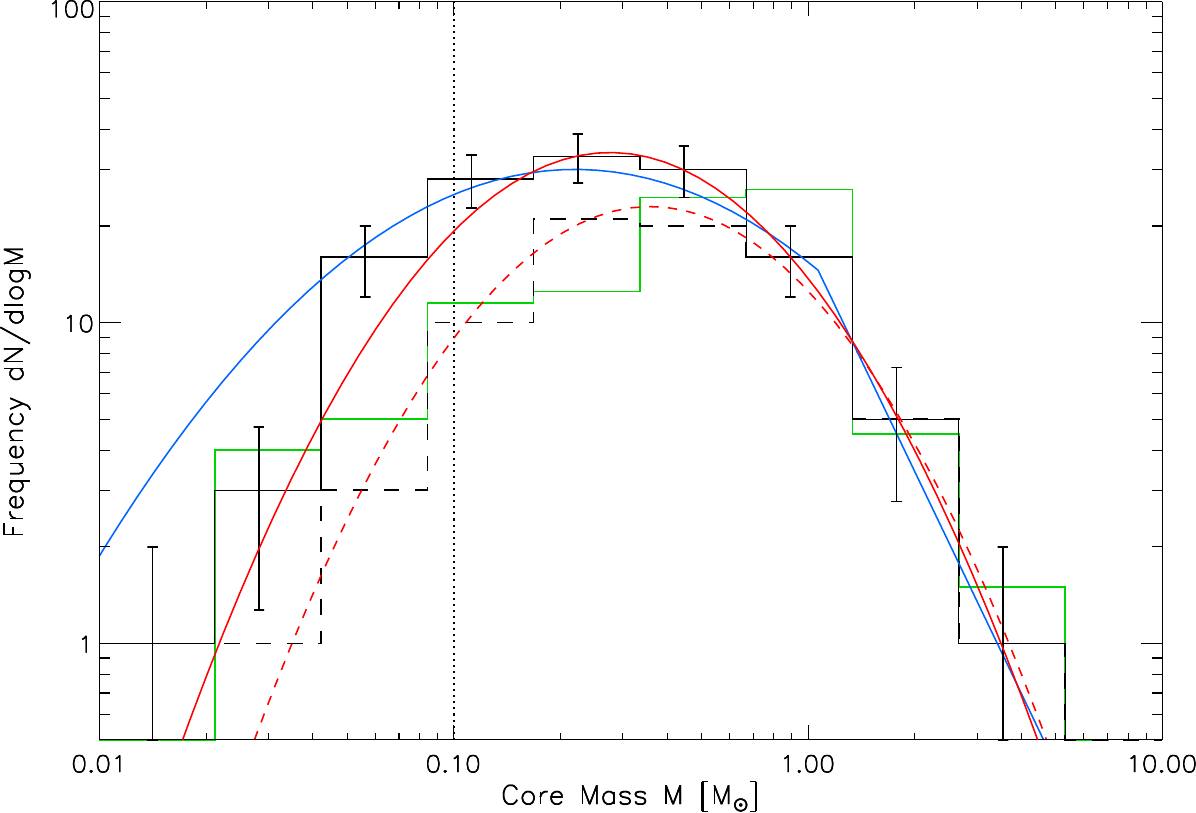} 
\caption{\label{fig:cmfcomp} The combined mass function for the L1495 and TMC1 regions. The solid black line shows all prestellar cores, the dashed black line shows robust prestellar cores. The red curve shows a log-normal fit to the prestellar core distribution above the completeness limit of this paper ($0.1$~M$_\odot$, shown by the vertical dotted-black line). The blue curve shows a scaled IMF from \citet{2003PASP..115..763C} and the green curve shows the Taurus IMF derived by \citet{2009ApJ...703..399L}. }
\end{figure}

Figure~\ref{fig:cmfcomp} shows a combined CMF for Taurus including robust and candidate prestellar cores from both TMC1 (this paper) and L1495 \citepalias{2016MNRAS.459..342M}. The error bars show $\sqrt{N}$ uncertainties. This curve is best fit by a lognormal curve that peaks at $0.28\pm0.08$~M$_\odot$. The curve has standard deviation of $0.42\pm0.06$, close to the value of 0.45 found for dense cores in Aquila \citep{2015A&A...584A..91K}. Also shown is the CMF for purely robust cores and an equivalent best fit lognormal curve. That fit peaks at  $0.36\pm0.08$~M$_\odot$ and has a standard deviation of $0.40\pm0.06$.

Figure~\ref{fig:cmfcomp} also shows a scaled version of the standard initial mass function from \citet{2003PASP..115..763C}. Note that the IMF for young stars in Taurus has been reported to differ from that seen in other regions, with an excess of stars between $0.7-1.0$~M$_\odot$ and a deficiency of stars above that mass \citep{2009ApJ...703..399L,2017ApJ...838..150K}. We therefore also show in Figure~\ref{fig:cmfcomp} the Taurus IMF from \citet{2009ApJ...703..399L}. This IMF has been scaled to match the number of cores at the high mass end. The Robust CMF most closely matches shape of the \citet{2009ApJ...703..399L} IMF as it peaks at a higher mass than the \citet{2003PASP..115..763C} IMF. Note, however, that neither the \citet{2003PASP..115..763C} nor the \citet{2009ApJ...703..399L} mass functions have been shifted horizontally. 

Scaled in this manner, there is an approximate 1-to-1 correlation between the Taurus CMF and IMF above a mass of $\sim0.3$~M$_\odot$, but a deficit in the CMF compared to the IMF below this. Although, given an uncertainty of $\sim2$ in core masses, assuming standard dust mass opacity uncertainties, a horizontal shift of the same magnitude between the CMF and IMF cannot be ruled out (c.f., sec 6. of \citetalias{2016MNRAS.459..342M}).

\section{Summary and Conclusions}

In this paper, we have used SPIRE and PACS parallel-mode maps to map and characterise the dense cores and filaments towards the TMC1 region in Taurus. Our principal conclusions can be summarised as follows:
\begin{enumerate}
\item The region mapped by this paper includes approximately $2000$~M$_\odot$ of material of which 34\% is above an extinction of $A_V\sim 3\ $mag. This approximate threshold appears as breaks in the cumulative histogram of column density and the column density PDF, and is the minimum background column density at which prestellar cores are found. Above this value, cores are shown to form more efficiently with increasing background density, reaching a peak around $A_V\sim14$.
\item A total of 35 robust filaments are identified following the methodology described in \citet{2019A&A...621A..42A}. The inner FWHM widths of these filaments are consistent with earlier studies. The supercritical TMC1 filaments are aligned orthogonal to the bulk magnetic field direction. However, there are relatively few of these overall and only $\sim10$\% of filaments have peak column densities in excess of an equivalent of $A_v~\approx7-8$.
\item The chemical young TMC1 filament is shown to be thermally supercritical. The South filament is largely undifferentiated. 
\item A catalogue of 44 robust and candidate prestellar cores is tabulated from across the TMC1 region that is 85 per cent complete above a core mass of $0.1$~M$_\odot$. The identified prestellar cores have a median temperature of 11.4~K and a median mass of 0.33~M$_\odot$. 
\item Star formation in the TMC1 region appears to be occurring at lower extinctions than other regions, with only 10\% of all filaments having an average extinction above 7--8 mag. 
\item The prestellar CMF for Taurus (L1495 and TMC1 regions combined) is well fit by a single log-normal distribution and is comparable with the standard IMF (e.g., \citealt{2003PASP..115..763C}). The prestellar CMF, however, does differ from the specific IMF derived for Taurus derived by \citet{2009ApJ...703..399L}.

\end{enumerate}

\section*{Acknowledgements}

SPIRE has been developed by a consortium of institutes led by Cardiff University (UK) and including Univ. Lethbridge (Canada); NAOC (China); CEA, LAM (France); IFSI, Univ. Padua (Italy); IAC (Spain); Stockholm Observatory (Sweden); Imperial College London, RAL, UCL-MSSL, UKATC, Univ. Sussex (UK); and Caltech, JPL, NHSC, Univ. Colorado (USA). This development has been supported by national funding agencies: CSA (Canada); NAOC (China); CEA, CNES, CNRS (France); ASI (Italy); MCINN (Spain); SNSB (Sweden); STFC, UKSA (UK); and NASA (USA).

PACS has been developed by a consortium of institutes led by MPE (Germany) and including UVIE (Austria); KU Leuven, CSL, IMEC (Belgium); CEA, LAM (France); MPIA (Germany); INAF-IFSI/OAA/OAP/OAT, LENS, SISSA (Italy); IAC (Spain). This development has been supported by the funding agencies BMVIT (Austria), ESA-PRODEX (Belgium), CEA/CNES (France), DLR (Germany), ASI/INAF (Italy), and CICYT/MCYT (Spain).

Based on observations obtained with Planck (\url{http://www.esa.int/Planck}), an ESA science mission with instruments and contributions directly funded by ESA Member States, NASA, and Canada.

The first author thanks J. Cernicharo and A. Hacar for useful discussions on column density comparisons. 

S.B. and N.S. acknowledge support by the french ANR and the german DFG through the project "GENESIS" (ANR‐16‐CE92‐0035‐01/DFG1591/2‐1).

\section*{Data Availability}

The data used in this paper, the full catalogue, and other HGBS data products, which include Herschel images and the column density and dust temperature maps are available from the HGBS website at \url{http://gouldbelt-herschel.cea.fr/archives}. 

\bibliographystyle{mnras}
\bibliography{tmc1}

\appendix

\section{Individual Wavelength Images}

\begin{figure}
\begin{center}
\includegraphics[width=\columnwidth]{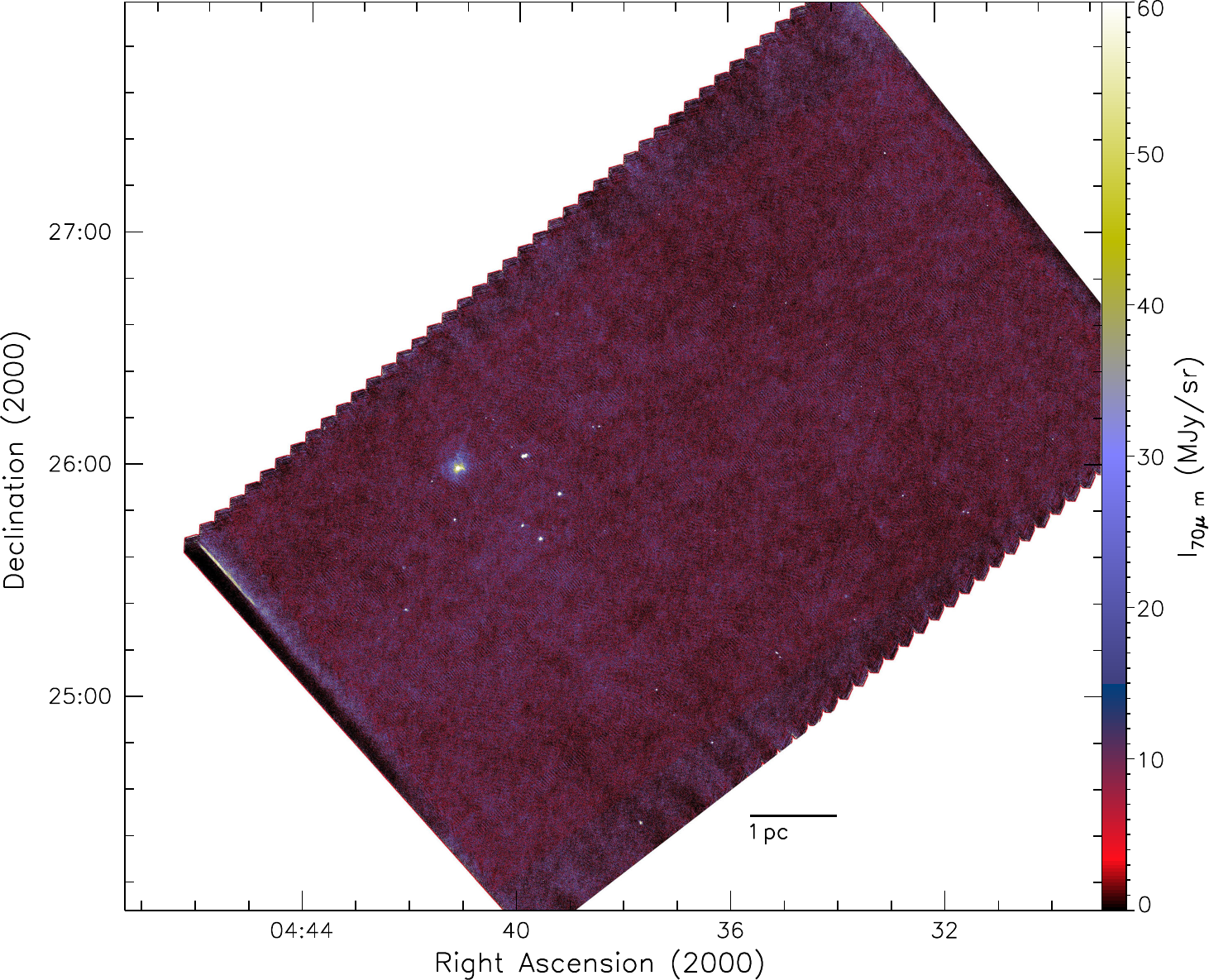}
\end{center}
\caption{\label{fig:flux070}\textit{Herschel} map of submillimetre dust emission at 70 $\mu$m towards the TMC1 region, as detected by \textit{Herschel}'s PACS instrument. The maps have been corrected for the IRAS/Planck DC offset (see Section~\ref{sec:structure}). The horizontal scale bar shows 1 pc at the assumed distance to Taurus (140 pc). }
\end{figure}

\begin{figure}
\begin{center}
\includegraphics[width=\columnwidth]{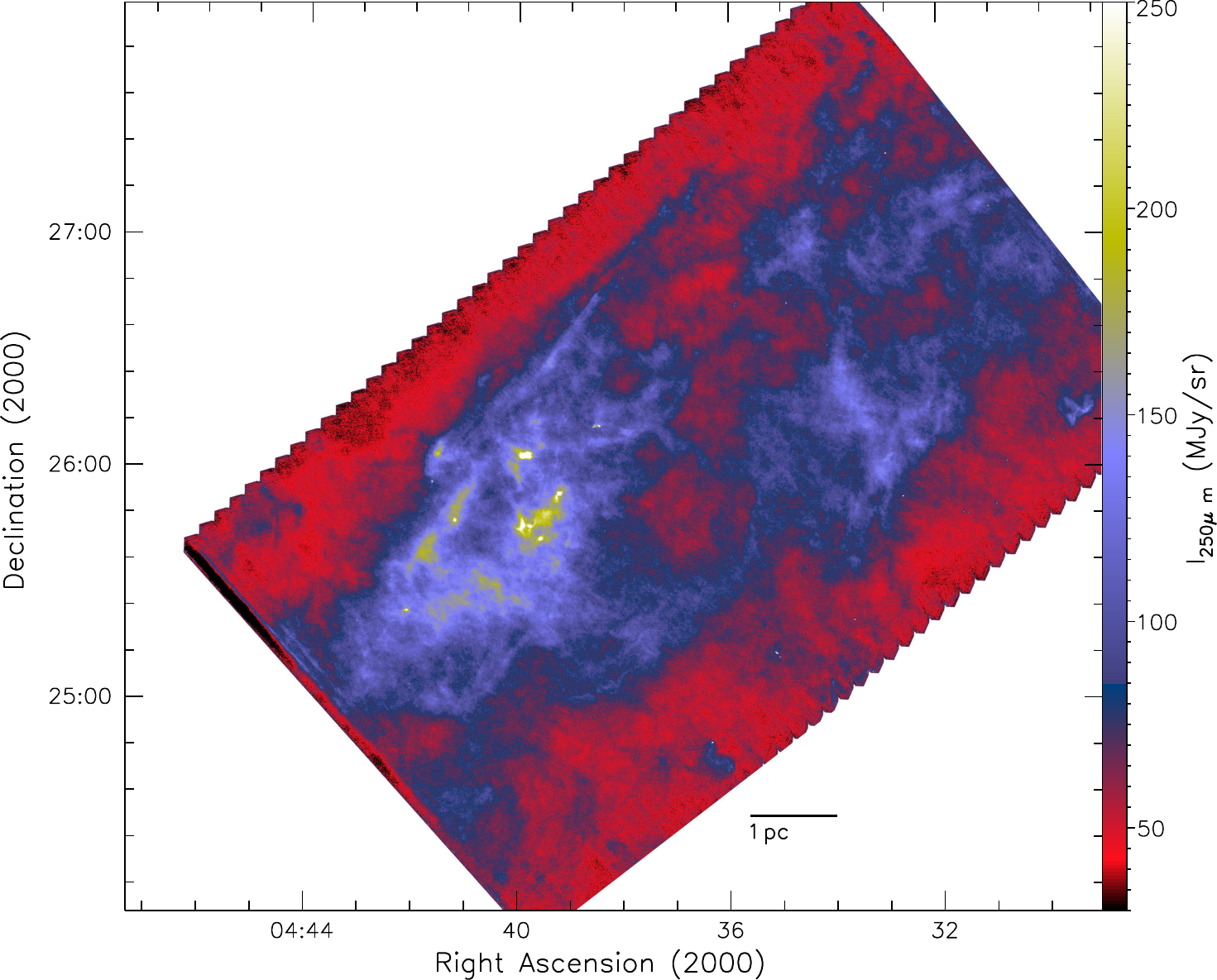}
\end{center}
\caption{\label{fig:flux160}\textit{Herschel} map of submillimetre dust emission at 160 $\mu$m towards the TMC1 region, as detected by \textit{Herschel}'s PACS instrument. Details as Figure~\ref{fig:flux070}. }
\end{figure}

\begin{figure}
\begin{center}
\includegraphics[width=\columnwidth]{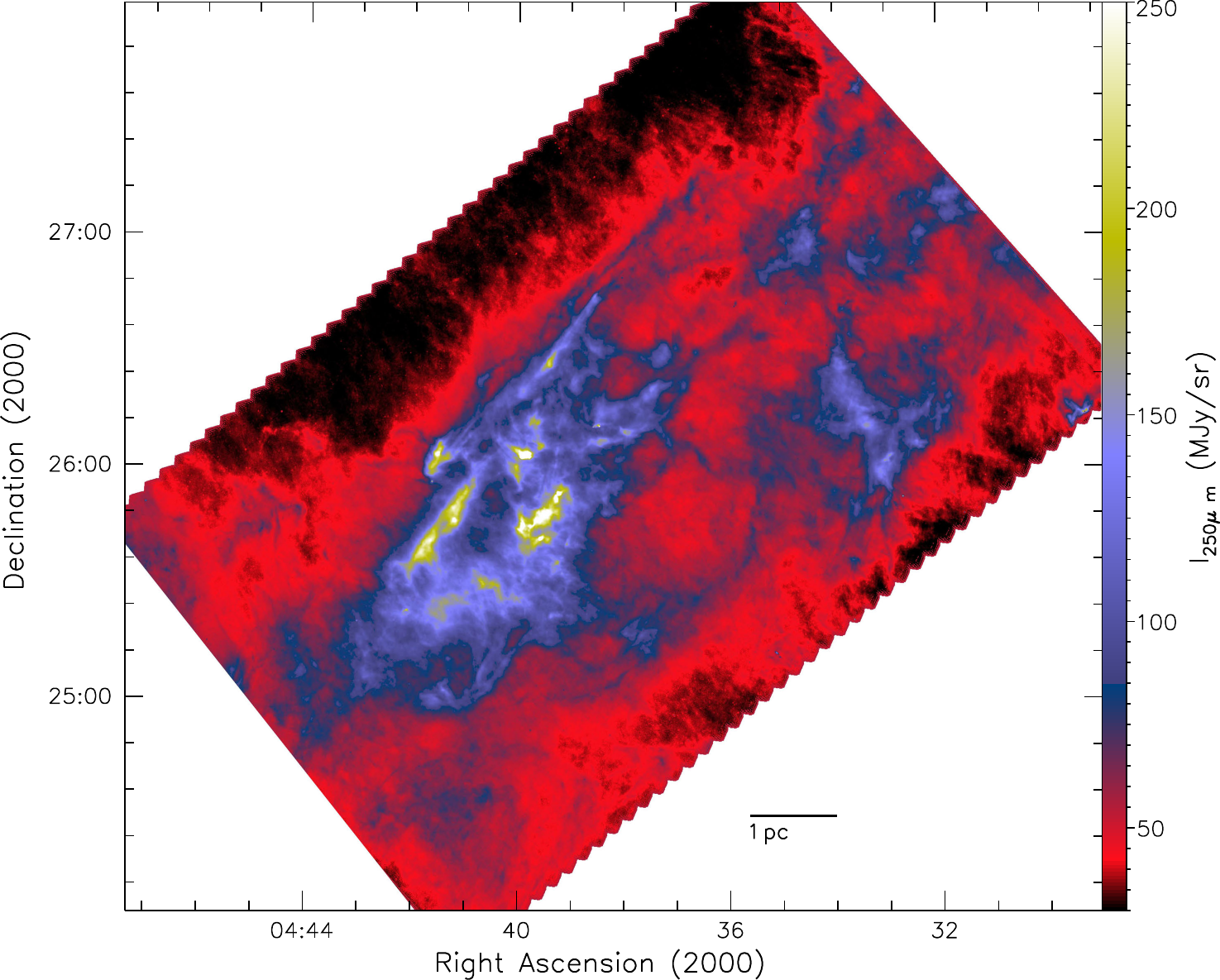}
\end{center}
\caption{\label{fig:flux250}\textit{Herschel} map of submillimetre dust emission at 250 $\mu$m towards the TMC1 region, as detected by \textit{Herschel}'s SPIRE instrument. Details as Figure~\ref{fig:flux070}.}
\end{figure}

\begin{figure}
\begin{center}
\includegraphics[width=\columnwidth]{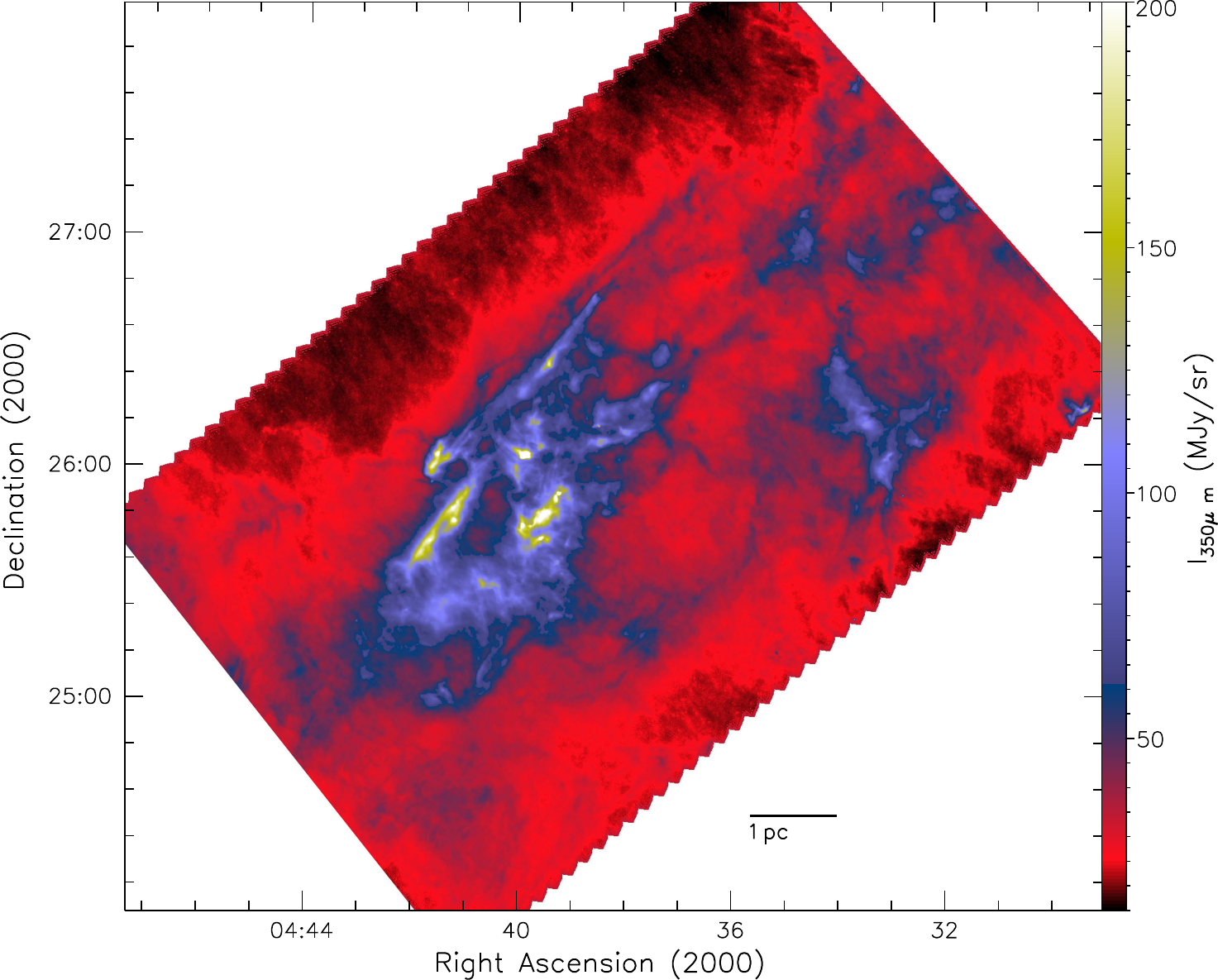}
\end{center}
\caption{\label{fig:flux350}\textit{Herschel} map of submillimetre dust emission at 350 $\mu$m towards the TMC1 region, as detected by \textit{Herschel}'s SPIRE instrument. Details as Figure~\ref{fig:flux070}. }
\end{figure}

\begin{figure}
\begin{center}
\includegraphics[width=\columnwidth]{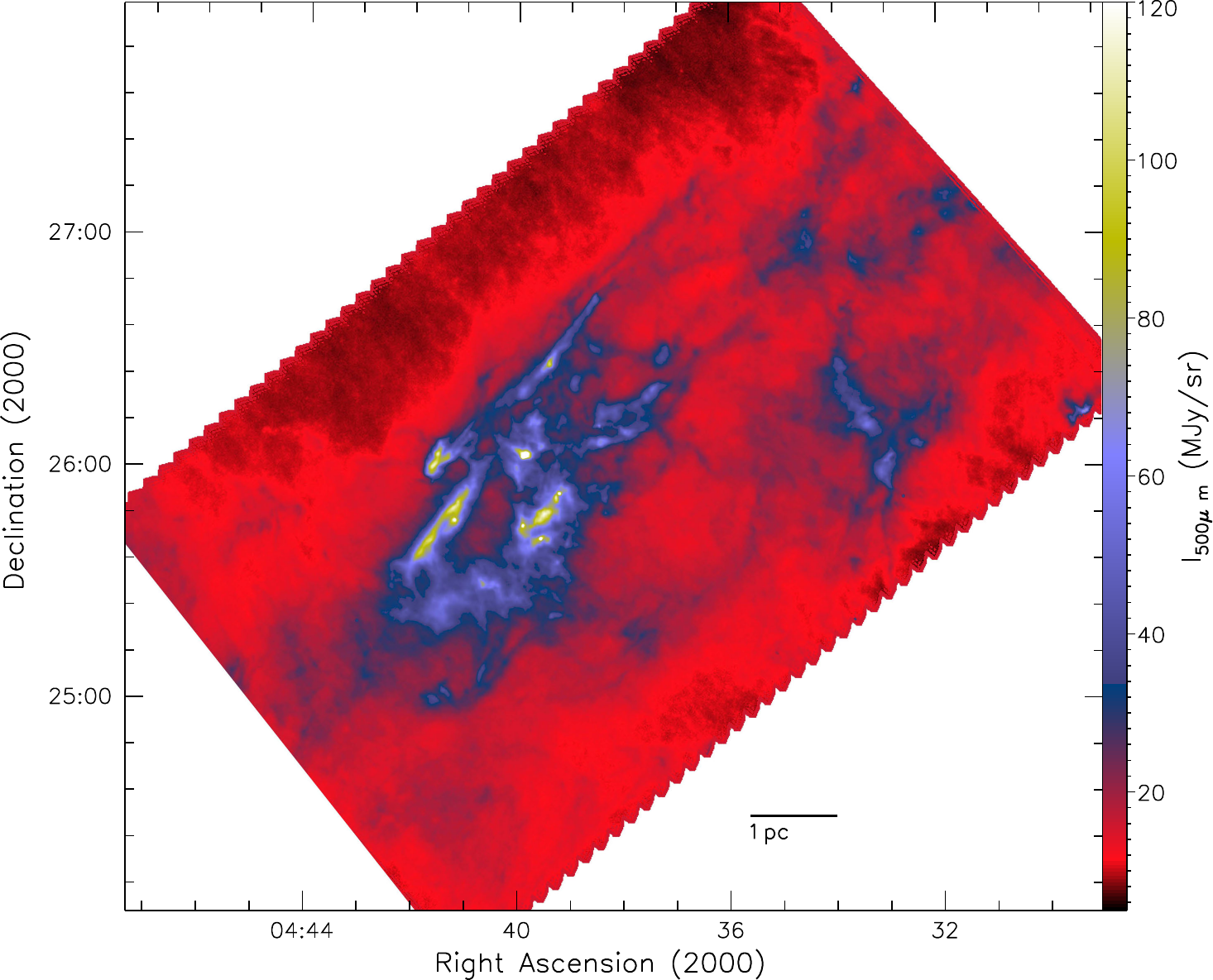}
\end{center}
\caption{\label{fig:flux500}\textit{Herschel} map of submillimetre dust emission at 500 $\mu$m towards the TMC1 region, as detected by \textit{Herschel}'s SPIRE instrument. Details as Figure~\ref{fig:flux070}.}
\end{figure}

Figures~\ref{fig:flux070} to \ref{fig:flux500} show the individual \textit{Herschel} flux maps for the TMC1 region. Details are listed in the caption to Figure~\ref{fig:flux070}. The data reduction process is described in Section~\ref{sec:obs}.  

\section{Dense Core Catalogue}

Table~\ref{tab:prop} shows the first three lines of the catalogue, the rest being available online. Each column of the table is numbered. Catalogue entries are as follows: \\
{\bf(1)} Core number; \\
{\bf(2)} Core name $=$ HGBS\_J prefix directly followed by a tag created from the J2000 sexagesimal coordinates; \\ 
{\bf(3)} and {\bf(4)}: Right ascension and declination of core center; \\
{\bf(5)}, {\bf(15)}, {\bf(25)}, {\bf(35)}, and {\bf(45)}: Detection significance from monochromatic single scales, in the 70-, 160-, 250-, 350-, and 500-$\mu$m maps, respectively. \\
{\bf(6)}$\pm${\bf(7)}, {\bf(16)}$\pm${\bf(17)} {\bf(26)}$\pm${\bf(27)} {\bf(36)}$\pm${\bf(37)} {\bf(46)}$\pm${\bf(47)}: Peak flux density and its error in Jy/beam as estimated by \textsl{getsources}; \\
{\bf(8)}, {\bf(18)}, {\bf(28)}, {\bf(38)}, {\bf(48)}: Contrast over the local background, defined as the ratio of the background-subtracted peak intensity to the local background intensity ($S^{\rm peak}_{\rm \lambda}$/$S_{\rm bg}$); \\
{\bf(9)}, {\bf(19)}, {\bf(29)}, {\bf(39)}: Peak flux density measured after smoothing to a 36.3$\arcsec$ beam; \\
{\bf(10)}$\pm${\bf(11)}, {\bf(20)}$\pm${\bf(21)}, {\bf(30)}$\pm${\bf(31)}, {\bf(40)}$\pm${\bf(41)}, {\bf(49)}$\pm${\bf(50)}: Integrated flux density and its error in Jy as estimated by \textsl{getsources}; \\
{\bf(12)}--{\bf(13)}, {\bf(22)}--{\bf(23)}, {\bf(32)}--{\bf(33)}, {\bf(42)}--{\bf(43)}, {\bf(51)}--{\bf(52)}: Major \& minor FWHM diameters of the core (in arcsec), respectively, as estimated by \textsl{getsources}. \\
{\bf(14)}, {\bf(24)}, {\bf(34)}, {\bf(44)}, {\bf(53)}: Position angle of the core major axis, measured east of north, in degrees; \\
{\bf(54)} Detection significance in the high-resolution column density image; \\
{\bf(55)} Peak H$_{2}$ column density in units of $10^{21}$ cm$^{-2}$ as estimated by \textsl{getsources} in the high-resolution column density image; \\
{\bf(56)} Column density contrast over the local background, as estimated by \textsl{getsources} in the high-resolution column density image;\\
{\bf(57)} Peak column density measured in a 36.3$\arcsec$ beam; \\
{\bf(58)} Local background H$_{2}$ column density as estimated by \textsl{getsources} in the high-resolution column density image; \\
{\bf(59)}--{\bf(60)}--{\bf(61)}: Major \& minor FWHM diameters of the core, and position angle of the major axis, respectively, as measured in the high-resolution column density image; \\
{\bf(62)} CSAR Counterpart: whether the \textsl{getsources} core has a counterpart detected by the \textsl{CSAR} source-finding algorithm: $1$ and $2$ denote when the \textsl{CSAR} counterpart is within an ellipse that is respectively equivalent to the half-width or full-width of the \textsl{getsources} core as measured on the high-resolution column density map. $3$ denotes if the \textsl{getsources} core is coincident with extended emission associated with a \textsl{CSAR} counterpart. \\
{\bf(63)} Number of \textit{Herschel} bands in which the core is significant (Sig$_{\rm \lambda} >$ 5) and has a positive flux density, excluding the column density plane; \\
{\bf(64)} Core type: either protostellar, robust prestellar, candidate prestellar, unbound starless, or other; \\

Table~\ref{tab_der_cat_cores} lists the derived properties of the dense cores, only the first three lines of the catalogue are shown, the rest being available online. Each column of the table is numbered. Entries are as follows: \\
 {\bf(1)} Core running number; {\bf(2)} Core name $=$ HGBS\_J prefix directly followed by a tag created from the J2000 sexagesimal coordinates; \\
{\bf(3)} and {\bf(4)}: Right ascension and declination of core center; \\
{\bf(5)} and {\bf(6)}: Geometrical average between the major and minor FWHM sizes of the core (in pc), as measured in the high-resolution column density map  before deconvolution, and after deconvolution from the 18.2$\arcsec$ HPBW resolution of the map, respectively. (NB: Both values provide estimates of the object's outer {\it radius} when the core can be approximately described by a Gaussian distribution, as is the case  for a critical Bonnor-Ebert spheroid); \\
{\bf(7)} Estimated core mass ($M_\odot$) assuming the dust opacity law advocated by \citet{2014A&A...562A.138R}; 
{\bf(9)} SED dust temperature (K); {\bf(8)} \& {\bf(10)} Statistical errors on the mass and temperature, respectively, including calibration uncertainties, but excluding dust opacity uncertainties; \\
{\bf(11)} Peak H$_2$ column density, at the resolution of the 500$~\mu$m data, derived from a graybody SED fit to the core peak flux densities measured in a common 36.3$\arcsec$ beam at all wavelengths; \\
{\bf(12)} Average column density, calculated as $N^{\rm ave}_{\rm H_2} = \frac{M_{\rm core}}{\pi R_{\rm core}^2} \frac{1}{\mu m_{\rm H}}$, 
          where $M_{\rm core}$ is the estimated core mass (col. {\bf 7}), $R_{\rm core}$ the estimated core radius prior to deconvolution (col. {\bf 6}), and $\mu = 2.86$;\\
{\bf(13)} Average column density calculated in the same way as for col. {\bf 12} but using the deconvolved core radius (col. {\bf 5}) instead of the core radius measured prior to deconvolution;  \\
{\bf(14)} Average volume density, calculated as\\
          $n^{\rm ave}_{\rm H_2} = \frac{M_{\rm core}}{4/3 \pi R_{\rm core}^3} \frac{1}{\mu m_{\rm H}}$, using the estimated core radius prior to deconvolution; \\
{\bf(15)} Average volume density, calculated in the same way as for col. {\bf 15} but using the deconvolved core radius (col. {\bf 5}) instead of the core radius measured prior to deconvolution; \\
{\bf(16)} Bonnor-Ebert mass ratio: $\alpha_{\rm BE} = M_{\rm BE,crit} / M_{\rm obs} $ (see text for details); \\
{\bf(17)} Core type: either protostellar, robust prestellar, candidate prestellar, unbound starless, or other;\\
{\bf(18)} Comments may be \textit{no SED fit}

\begin{table*}\setlength{\tabcolsep}{2.5pt}
\caption{Catalogue of dense cores identified in the HGBS maps of the TMC1 Region of the Taurus molecular cloud (template, full catalog only provided online). The header's third row contains column references explained in the main text. } 
\label{tab:prop}
\renewcommand{\arraystretch}{1.2}
\begin{tabular}{ | c | c | r@{:}c@{:}l r@{:}c@{:}l | c r l c c l r c c c }
\hline\hline 
 C.No. & Source name & \multicolumn{3}{c}{R.A. (2000)} & \multicolumn{3}{c}{Dec. (2000)} & Sig$_{70}$ & \multicolumn{2}{c}{$S^\mathrm{peak}_{70}$} & $S^\mathrm{peak}_{70}/S_\mathrm{bg}$ & $S^{\textrm{conv,}500}_{70}$ & \multicolumn{2}{c}{$S^\mathrm{tot}_{70}$} & a$_{70}$ & b$_{70}$ & PA$_{70}$  \\ 
 & & \multicolumn{3}{c}{(h m s)} & \multicolumn{3}{c}{($^{\circ}$~\arcmin~\arcsec)} &  & \multicolumn{2}{c}{(Jy~beam$^{-1}$)} &  & (Jy~beam$^{-1}_{500}$) & \multicolumn{2}{c}{(Jy)} & (\arcsec) & (\arcsec) & ($^{\circ}$) \\ 
 (1) & (2) & \multicolumn{3}{c}{(3)} & \multicolumn{3}{c}{(4)} & (5) & \multicolumn{1}{c}{(6)} & \multicolumn{1}{c}{(7)} & (8) & (9) & \multicolumn{1}{c}{(10)} & \multicolumn{1}{c}{(11)} & (12) & (13) & (14) \\ 
\hline
     6 & 043222.7+270051 & 04 & 32 & 22.75 & +27 & 00 & 51.5 & 1.3 & -5.13e-03 & 4.0e-03 & 0.72 & -1.05e-01 & -1.83e-01 & 1.4e-01 & 234 & 93 & 119 \\
     9 & 043242.9+255231 & 04 & 32 & 42.95 & +25 & 52 & 31.2 & 93.2 & 1.48e+00 & 1.7e-02 & 227.83 & 1.75e+00 & 2.05e+00 & 2.4e-02 & 8 & 8 & 198 \\
     14 & 043307.9+255848 & 04 & 33 & 07.99 & +25 & 58 & 48.5 & 0.0 & 9.65e-03 & 6.0e-03 & 0.88 & 2.05e-01 & 4.85e-01 & 3.0e-01 & 178 & 170 & 52 \\
         
\hline\hline
\end{tabular}
\scalebox{1.2}{$\sim$}
\vspace{0.2cm}

\scalebox{1.2}{$\sim$}
{\renewcommand{\arraystretch}{1.2}
\begin{tabular}{ c r l c c l r c c c c r l c   }
\hline\hline 
  Sig$_{160}$ & \multicolumn{2}{c}{$S^\mathrm{peak}_{160}$} & $S^\mathrm{peak}_{160}/S_\mathrm{bg}$ & $S^{\textrm{conv,}500}_{160}$ & \multicolumn{2}{c}{$S^\mathrm{tot}_{160}$} & a$_{160}$ & b$_{160}$ & PA$_{160}$ &   Sig$_{250}$ & \multicolumn{2}{c}{$S^\mathrm{peak}_{250}$} & $S^\mathrm{peak}_{250}/S_\mathrm{bg}$   \\ 
  & \multicolumn{2}{c}{(Jy~beam$^{-1}$)} &  & (Jy~beam$^{-1}_{500}$) & \multicolumn{2}{c}{(Jy)} & (\arcsec) & (\arcsec) & ($^{\circ}$) &  & \multicolumn{2}{c}{(Jy~beam$^{-1}$)} &    \\
 (15) & \multicolumn{1}{c}{(16)} & \multicolumn{1}{c}{(17)} & (18) & (19) & \multicolumn{1}{c}{(20)} & \multicolumn{1}{c}{(21)} & (22) & (23) & (24) & (25) & \multicolumn{1}{c}{(26)} & \multicolumn{1}{c}{(27)} & (28)   \\
\hline
   0.0 & 9.89e-02 & 3.0e-02 & 0.50 & 8.26e-01 & 2.26e+01 & 6.9e+00 & 226 & 192 & 161 & 8.1 & 3.75e-01 & 3.6e-02 & 2.57   \\	
   86.5 & 2.21e+00 & 2.6e-02 & 45.10 & 2.62e+00 & 2.03e+00 & 2.4e-02 & 13 & 13 & 117 & 109.8 & 1.64e+00 & 3.5e-02 & 34.78  \\	
  0.0 & 3.09e-02 & 2.0e-02 & 0.10 & 2.56e-01 & 8.84e+00 & 5.7e+00 & 194 & 178 & 189 & 7.9 & 2.18e-01 & 3.0e-02 & 0.36   \\
  \hline\hline
\end{tabular}
}
\scalebox{1.2}{$\sim$}
\vspace{0.2cm}

\scalebox{1.2}{$\sim$}
{\renewcommand{\arraystretch}{1.2}
\begin{tabular}{ c l r c c c c r l c c l r c c c c   }
\hline\hline 
   $S^{\textrm{conv,}500}_{250}$ & 
   \multicolumn{2}{c}{$S^\mathrm{tot}_{250}$} & 
   a$_{250}$ & b$_{250}$ & PA$_{250}$ & 
   Sig$_{350}$ & 
   \multicolumn{2}{c}{$S^\mathrm{peak}_{350}$} & 
   $S^\mathrm{peak}_{350}/S_\mathrm{bg}$ & 
   $S^{\textrm{conv,}500}_{350}$ & 
   \multicolumn{2}{c}{$S^\mathrm{tot}_{350}$} & 
   a$_{350}$ & b$_{350}$    \\ 
   
  (Jy~beam$^{-1}_{500}$) & 
  \multicolumn{2}{c}{(Jy)} & 
  (\arcsec) & (\arcsec) & ($^{\circ}$) & & 
  \multicolumn{2}{c}{(Jy~beam$^{-1}$)} &  & (Jy~beam$^{-1}_{500}$) & \multicolumn{2}{c}{(Jy)} & (\arcsec) & (\arcsec)  \\
   (29) & \multicolumn{1}{c}{(30)} & \multicolumn{1}{c}{(31)} & (32) & (33) & (34) & (35) & \multicolumn{1}{c}{(36)} & \multicolumn{1}{c}{(37)} & (38) & (39) & \multicolumn{1}{c}{(40)} & \multicolumn{1}{c}{(41)} & (42) & (43)   \\
\hline
     1.47e+00 & 2.56e+01 & 2.4e+00 & 135 & 131 & 125 & 9.4 & 3.80e-01 & 5.0e-02 & 2.03 & 8.29e-01 & 1.13e+01 & 1.5e+00 & 126 & 108   \\
     1.71e+00 & 1.36e+00 & 2.9e-02 & 18 & 18 & 54 & 68.3 & 1.31e+00 & 5.5e-02 & 16.67 & 1.41e+00 & 1.10e+00 & 4.6e-02 & 24 & 24   \\
     8.33e-01  & 8.82e+00 & 1.2e+00 & 125 & 95 & 202 & 11.2 & 4.76e-01 & 1.1e-01 & 0.66 & 1.03e+00 & 1.06e+01 & 2.5e+00 & 119 & 92   \\	     
\hline\hline
\end{tabular}
}
\scalebox{1.2}{$\sim$}
\vspace{0.2cm}

\scalebox{1.2}{$\sim$}
{\renewcommand{\arraystretch}{1.2}
\begin{tabular}{ r l c c l r c c c c c c c c c c c c  }
\hline\hline 
  PA$_{350}$ & Sig$_{500}$ & \multicolumn{2}{c}{$S^\mathrm{peak}_{500}$} & $S^\mathrm{peak}_{500}/S_\mathrm{bg}$ & \multicolumn{2}{c}{$S^\mathrm{tot}_{500}$} & a$_{500}$ & b$_{500}$ & PA$_{500}$ & Sig$_{\textrm{N}(\textrm{H}_{2})}$ & $N^\mathrm{peak}_{\textrm{H}_{2}}$ & $N^\mathrm{peak}_{\textrm{H}_{2}}/N_\mathrm{bg}$ & $N^{\textrm{conv,}500}_{\textrm{H}_{2}}$  \\

  ($^{\circ}$) & & \multicolumn{2}{c}{(Jy~beam$^{-1}$)} & &  \multicolumn{2}{c}{(Jy)} & (\arcsec) & (\arcsec) & ($^{\circ}$) & & (10$^{21}$~cm$^{-2}$) &   & (10$^{21}$~cm$^{-2}$)  \\
  (44) & (45) & \multicolumn{1}{c}{(46)} & \multicolumn{1}{c}{(47)} & (48) & \multicolumn{1}{c}{(49)} & \multicolumn{1}{c}{(50)} & (51) & (52) & (53) & (54) & (55) & (56) & (57)  \\
  \hline
    110 & 12.3 & 4.22e-01 & 4.1e-02 & 2.44 & 6.29e+00 & 6.1e-01 & 127 & 113 & 119 & 13.7 & 1.77e+00 & 0.88 & 1.65e+00  \\
    152 & 55.1 & 8.82e-01 & 3.8e-02 & 8.06 & 7.13e-01 & 3.0e-02 & 36 & 36 & 180 & 69.3 & 4.66e+00 & 2.34 & 1.31e+00 \\
    195 & 17.4 & 5.11e-01 & 6.9e-02 & 0.64 & 6.30e+00 & 8.5e-01 & 123 & 109 & 202 & 16.9 & 1.70e+00 & 0.40 & 1.51e+00 \\

\hline\hline
\end{tabular}
\scalebox{1.2}{$\sim$}
\vspace{0.2cm}
}

\scalebox{1.2}{$\sim$}
{\renewcommand{\arraystretch}{1.2}
\begin{tabular}{ c c c c c c c | }
\hline\hline 
   $N^{\textrm{bg}}_{\textrm{H}_{2}}$ & a$_{\textrm{N}(\textrm{H}_{2})}$ & b$_{\textrm{N}(\textrm{H}_{2})}$ & PA$_{\textrm{N}(\textrm{H}_{2})}$ &   \textsc{csar} & N$_{\textrm{SED}}$  \\

  (10$^{21}$~cm$^{-2}$) & (\arcsec) & (\arcsec) & ($^{\circ}$) & &  \\
   (58) & (59) & (60) & (61) & (62) & (63) & (64)  \\
  \hline
      2.00e+00 & 133 & 102 & 112 & 1 & 3 & unbound starless \\
      2.00e+00 & 21 & 18 & 170 & 1 & 5 & protostellar \\
      4.22e+00 & 101 & 81 & 187 & 1 & 3 & robust prestellar \\

\hline\hline
\end{tabular}
}

\end{table*}

\begin{table*}\setlength{\tabcolsep}{4.0pt}
\caption{Derived properties of the dense cores identified in the HGBS maps of the TMC1 Region of the Taurus molecular cloud (template, full table only provided online). The header's third row contains column references explained in the main text.}
\label{tab_der_cat_cores}

{\renewcommand{\arraystretch}{1.2}
\begin{tabular}{ | c | c | r@{:}c@{:}l r@{:}c@{:}l | c c c c c c c c c c c |}
\hline\hline 
 C.No. & Source name & \multicolumn{3}{c}{R.A. (2000)} & \multicolumn{3}{c}{Dec. (2000)} & \multicolumn{2}{c}{$R_{\textrm{core}}$} & \multicolumn{2}{c}{$M_{\textrm{core}}$} & \multicolumn{2}{c}{$T_{\textrm{core}}$} & $N^\mathrm{\textrm{peak}}_{\textrm{H}_{2}}$ & \multicolumn{2}{c}{$N^\mathrm{ave}_{\textrm{H}_{2}}$}  \\ 
 & & \multicolumn{3}{c}{(h m s)} & \multicolumn{3}{c}{($^{\circ}$~\arcmin~\arcsec)} & \multicolumn{2}{c}{(pc)} & \multicolumn{2}{c}{(M$_{\odot}$)} & \multicolumn{2}{c}{(K)} & (10$^{21}$~cm$^{-2}$) & \multicolumn{2}{c}{(10$^{21}$~cm$^{-2}$)} \\ 
 (1) & (2) & \multicolumn{3}{c}{(3)} & \multicolumn{3}{c}{(4)} & (5) & \multicolumn{1}{c}{(6)} & \multicolumn{1}{c}{(7)} & (8) & (9) & \multicolumn{1}{c}{(10)} & \multicolumn{1}{c}{(11)} & (12) & (13) \\ 
\hline

     6 & 043222.7+270051 & 04 & 32 & 22.75 & +27 & 00 & 51.5 & 0.079 & 0.078 & 0.276 & 0.081 & 14.9 & 1.3 & 1.80e+21 & 6.1e-01 & 6.3e-01 \\
     9 & 043242.9+255231 & 04 & 32 & 42.95 & +25 & 52 & 31.2 & 0.013 & 0.005 & 0.024 & 0.005 & 14.4 & 0.6 & 1.64e+21 & 1.9e+00 & 1.3e+01  \\
    14 & 043307.9+255848 & 04 & 33 & 07.99 & +25 & 58 & 48.5 & 0.062 & 0.060 & 1.035 & 0.416 & 9.5 & 0.8 & 4.93e+21 & 3.8e+00 & 4.0e+00  \\

\hline\hline
\end{tabular}
\scalebox{1.2}{$\sim$}
\vspace{0.2cm}

\scalebox{1.2}{$\sim$}
\begin{tabular}{ |c c c c c|}
\hline\hline 
 \multicolumn{2}{c}{$n^\mathrm{ave}_{\textrm{H}_{2}}$} & $\alpha_{\textrm{BE}}$ & Core type & comments \\ 
\multicolumn{2}{c}{(10$^{4}$~cm$^{-3}$)} & & & \\ 
 (14) &(15)&(16)&(17) & (18) \\ 
\hline

     1.9e-01 & 1.9e-01 & 8.2 & unbound starless & \\
     3.4e+00 & 6.3e+01 & 5.9 & protostellar & \\
     1.5e+00 & 1.6e+00 & 1.1 & robust prestellar & \\

\hline\hline
\end{tabular}
}

\end{table*}

\bsp	
\label{lastpage}
\end{document}